\documentclass[pra,aps,floatfix,amsmath,superscriptaddress,twocolumn,longbibliography,nofootinbib]{revtex4-2}
\usepackage{amssymb,enumerate}

\usepackage{amssymb,enumerate}
\usepackage{graphicx}
\usepackage{graphics}
\usepackage{amsmath}
\usepackage{amsthm,bbm}
\usepackage{color}
\usepackage{dsfont}
\usepackage{hyperref}

\usepackage[capitalise,nameinlink]{cleveref}

\usepackage{qcircuit}

\usepackage{graphicx}

\usepackage{xcolor}
\hypersetup{
    colorlinks,
    linkcolor={red!50!black},
    citecolor={blue!50!black},
    urlcolor={blue!80!black}
}

\usepackage{xfrac}

\usepackage{amsmath}
\usepackage{tikz-cd}
\usepackage{empheq}

\usepackage{enumitem}

\usepackage{graphicx}

\usepackage{hyperref}
\usepackage[capitalise]{cleveref}
\usepackage{youngtab}

\usepackage{tikz}
\usepackage{mathtools}

\renewcommand{\i}{\mathrm{i}}

\newcommand{\Tr}{\operatorname{Tr}}

\newcommand{\<}{\langle}
\renewcommand{\>}{\rangle}

\renewcommand{\i}{\mathrm{i}}

\renewcommand{\i}{\mathrm{i}}

\newtheorem{thm}{Theorem}

\usepackage{lipsum}
\usepackage{lmodern}
\usepackage{tcolorbox}
\usepackage{orcidlink}

\usepackage{amsfonts}
\usepackage{graphicx,graphics,epsfig,times,bm,bbm,amssymb,amsmath,amsfonts,mathrsfs}
\usepackage[normalem]{ulem}
\usepackage{setspace}
\usepackage{subfigure}
\usepackage{dsfont}
\usepackage{braket}
\usepackage{upgreek }
\usepackage{tikz}
\usepackage{natbib}
\usepackage{chngcntr}

\newtheorem{theorem}{Theorem}

\newtheorem{lemma}[theorem]{Lemma}

\makeatletter 
\renewcommand\onecolumngrid{%
  \do@columngrid{one}{\@ne}%
  \def\set@footnotewidth{\onecolumngrid}%
  \def\footnoterule{\kern-6pt\hrule width 1.5in\kern6pt}%
}
\makeatother

\newcommand{\bes} {\begin{subequations}}
\newcommand{\ees} {\end{subequations}}
\newcommand{\bea} {\begin{eqnarray}}
\newcommand{\eea} {\end{eqnarray}}
\newcommand{\be} {\begin{equation}}
\newcommand{\ee} {\end{equation}}

\def\>{\rangle}
\def\<{\langle}
\def\Tr{\textrm{Tr}}

\newcommand{\ignore}[1]{}

\crefformat{step}{#2Step #1#3}

\begin{document}


	\title{Observation of the symmetry-protected signature of  3-body interactions}

\author{Liudmila A. Zhukas\orcidlink{0000-0002-1556-0081}}
\email{Corresponding author: liudmila.zhukas@duke.edu}
\affiliation{Duke Quantum Center, Departments of Electrical and Computer Engineering and Physics, Duke University, Durham, NC 27708}

\author{Qingfeng Wang\orcidlink{0000-0002-6199-1560}}
\affiliation{Duke Quantum Center, Departments of Electrical and Computer Engineering and Physics, Duke University, Durham, NC 27708}
\affiliation{%
  Chemical Physics Program and Institute for Physical Science and Technology, University of Maryland, College Park, Maryland, USA}

\author{Or Katz\orcidlink{0000-0001-7634-1993}}
\affiliation{Duke Quantum Center, Departments of Electrical and Computer Engineering and Physics, Duke University, Durham, NC 27708}
\affiliation{School of Applied and Engineering Physics, Cornell University, Ithaca, NY 14853.}

\author{Christopher Monroe}
\affiliation{Duke Quantum Center, Departments of Electrical and Computer Engineering and Physics, Duke University, Durham, NC 27708}

\author{Iman Marvian}
\affiliation{Duke Quantum Center, Departments of Electrical and Computer Engineering and Physics, Duke University, Durham, NC 27708}

\begin{abstract}
Identifying and characterizing multi-body interactions in quantum processes remains a significant challenge. This is partly because 2-body interactions can produce an arbitrary time evolution, a fundamental fact often called the universality of 2-local gates in the context of quantum computing. However, when an unknown Hamiltonian respects a U(1) symmetry such as charge or particle number conservation, N-body interactions exhibit a distinct symmetry-protected signature known as the N-body phase, which fewer-body interactions cannot mimic. We develop and demonstrate an efficient technique for the detection of 3-body interactions despite the presence of unknown 2-body interactions. This technique, which takes advantage of GHZ states for phase estimation, requires probing the unitary evolution and measuring its determinant in a small subspace that scales linearly with the system size, making it an efficient approach. 

\end{abstract}

\maketitle

\color{black}


 
 Recent developments in quantum information science have revealed various new possibilities for generating and controlling multi-body interactions \cite{katz2023programmable, katz2023demonstration, marvian2022restrictions2, alhambra2022forbidden, marvian2024rotationally}. However, while such interactions can be a powerful resource in quantum computing, characterizing and identifying them also pose new challenges for standard approaches in this field. Therefore, developing new methods for detecting and learning such interactions is a timely and important goal for the field of quantum metrology and Hamiltonian learning \cite{GiovannettiLloydScience, giovannetti2011advances, degen2017quantum, huang2023learning}. Multi-body interactions appear in natural physical processes and may exist fundamentally, as seen in quantum chromodynamics, or emerge within effective theories when certain degrees of freedom or subspaces are integrated out  \cite{hammer2013colloquium}. Given the successful use of quantum metrology techniques for enhancing sensor signal-to-noise such as the detection of gravitational waves in the LIGO/VIRGO experiments \cite{tse2019quantum, acernese2019increasing}, it is natural to consider the use of such techniques for detecting other elusive physical phenomena such as 3-body interactions. \\
 
\begin{figure}
{\includegraphics[scale=.45]{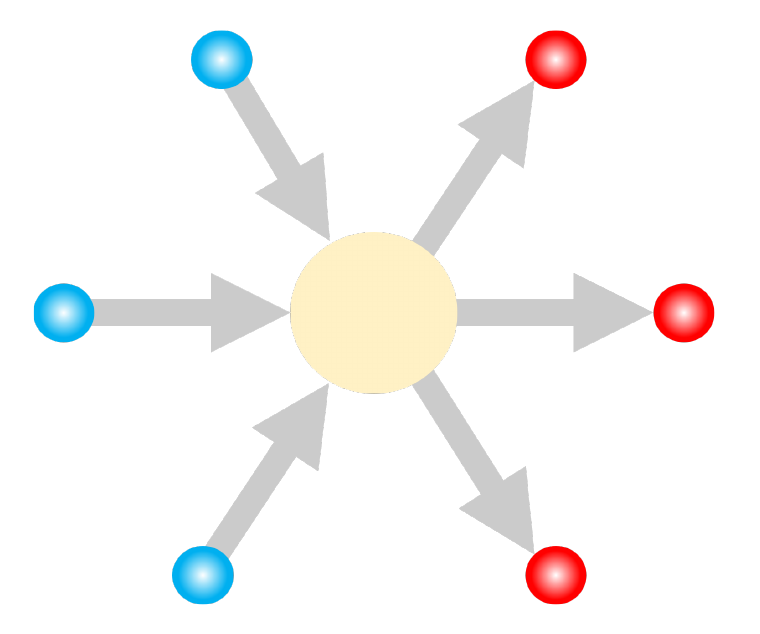}}\caption{\textbf{3-body interactions:} The Hamiltonian in Eq.(\ref{3local}) can be taken as a description of the interaction between the (blue) internal degrees of freedom of particles in a scattering process.    
By observing the (red) outputs of this process for different inputs, can we detect the presence of 3-body interactions $\sum a^{(3)}_{ijk}(t)\ Z_iZ_{j} Z_{k}$?}  
 \label{Fig:scattering}
\end{figure}

\noindent\textbf{A no-go theorem--} Do dynamics under 3-body interactions have any distinct signatures that cannot be reproduced by 2-body interactions, making them directly detectable? As an example, consider a system with $n$ qubits  interacting under a general Hamiltonian 
\be\label{3local}
H^{(3)}(t)=H^{(2)}(t)+\sum_{i<j<k} a^{(3)}_{ijk}(t)\ Z_iZ_{j} Z_{k}\ \ \ \  \ \ \ : 0\le t\le T\ ,
\ee
where $H^{(2)}(t)$ can be decomposed as a sum of 2-local interactions, i.e., those that act non-trivially only on, at most, a pair of qubits, and  $Z_i$ denote Pauli $Z$ operator on qubit $i$. As a concrete example, one may assume  $H^{(2)}(t)$ contains a time-dependent  XY interaction  $X\otimes X+Y\otimes Y$ between nearest-neighbor qubits on a chain, as well as time-dependent single-qubit $X$ and $Z$ fields. The Hamiltonian $H^{(3)}$ may describe, e.g.,  a spin chain or the interaction between the internal degrees of freedom of particles that interact through a scattering process (See Fig.~\ref{Fig:scattering}).



Suppose we can prepare the system in arbitrary initial states which then evolve under the Hamiltonian $H^{(3)}(t)$ from time $t=0$ to $T$.  By performing such experiments can we 
obtain any information about the hypothetical 3-body term $\sum a^{(3)}_{ijk}(t)\ Z_iZ_{j} Z_{k}$? It turns out that the answer is negative. \\

\noindent\textbf{Theorem:} 
Consider a unitary process described by an unknown Hamiltonian $H^{(3)}(t)$ in the form of Eq.(\ref{3local}) for a fixed total time $T$. Suppose one can prepare arbitrary initial states and perform arbitrary measurements on the final state, and repeat this arbitrarily many times.  Unless one has further information about the term $H^{(2)}(t)$ corresponding to the 2-body interactions in this Hamiltonian, it is impossible to detect the presence of 3-body interactions $\sum a^{(3)}_{ijk}(t)\ Z_iZ_{j} Z_{k}$, or to obtain any information about their strengths $a^{(3)}_{ijk}(t)$  .\\

This no-go theorem is an immediate corollary of the universality of 2-qubit gates, which is one of the cornerstones of quantum computing \cite{divincenzo1995two, lloyd1995almost}: By choosing different input states and output measurements,  the only information that can be obtained about the Hamiltonian $H^{(3)}$ is limited to what can be inferred from the unitary operator $V=\mathcal{T}\big\{\exp\big( {-\i \int_{0}^T H^{(3)}(t) dt}\big) \big\}$,   where $\mathcal{T}$ denotes the time-ordered integral and we take $\hbar=1$.  However, 2-body interactions in the Hamiltonian $H^{(3)}$ can synthesize any arbitrary unitary transformation. Therefore, adding three-body interactions does not extend the set of realizable unitaries, and thus their presence cannot be detected.


On the other hand, perhaps surprisingly, it has been recently shown that in the presence of symmetries, this no-go theorem can be violated \cite{marvian2022restrictions2}.  That is, 3-body interactions have certain signatures 
that are protected by the symmetry of the time evolution
and cannot be reproduced by 2-body interactions. 
Based on this observation, Ref.\cite{marvian2022restrictions2} proposes a method for detecting the locality of interactions. However, as explained below, this method requires probing the unknown unitary $V$ in the entire Hilbert space, rendering it inefficient for large systems.  

In this article, we introduce a new symmetry-protected signature of 3-body interactions, denoted by the phase $\Delta_3$, and develop a method for efficient measurement of this quantity. Furthermore, we perform an experiment that measures this quantity on trapped atomic ion qubits, directly detecting the presence of 3-body interactions. A key component of our scheme, which enables efficient detection of 3-body interactions is the use of Greenberger–Horne–Zeilinger (GHZ) states \cite{greenberger1989going}. Hence, our work reveals a novel application of GHZ states in the context of quantum sensing and Hamiltonian learning. \\

\section*{Main results}

\noindent \textbf{An observable phase protected by symmetry--}  Consider $n$ qubits evolving under Hamiltonian $H(t)$ from time $t=0$ to $t=T$. Assume the Hamiltonian respects the U(1) symmetry corresponding to rotations around the $z$-axis, or equivalently, it commutes with $\sum_j Z_j$. The  unitary time evolution $V$ generated under this Hamiltonian can then be decomposed as
\be
 V=\mathcal{T}\big\{\exp\big( {-\i \int_{0}^T H(t) dt}\big) \big\}=\bigoplus_{m=0}^n V_m\ ,
 \ee
where $V_m$ is the component of the unitary 
$V$ in the subspace with $m$ ``excitations'', or the eigen-subspace of $\sum_j (\mathbb{I}-Z_j)/2$ with eigenvalue $m$ (also, known as the subspace with the Hamming weight $m$). The U(1) symmetry implies that the number of excitations $m$ is conserved under unitary $V$, which explains the above block-diagonal form.  For any such unitary, we define the phase
\begin{align}\label{Phase}
\Delta_3\equiv \theta_{n-1}-\theta_{1}-(n-2)\times (\theta_n-\theta_0) \pmod {2\pi},
\end{align}
where   $$\theta_m\equiv\text{arg}[\text{det}(V_m)]\in (-\pi,\pi]$$ 
is the phase of the determinant of unitary $V_m$. Note that since $m=0, n$ sectors are one-dimensional,  $\theta_0=\text{arg}(V_0)$ and  $\theta_n=\text{arg}(V_n)$. 
 
 As we prove in Methods, the phase $\Delta_3$ has the following properties: 
First,  if the U(1)-invariant Hamiltonian $H(t)$ only contains 2-qubit interactions, then $\Delta_3$ is an integer multiple of $2\pi$. Therefore,  $\Delta_3\neq 0 \pmod{2\pi}$ indicates the presence of interactions that couple more than 2 qubits together.   In particular, for Hamiltonian $H^{(3)}$ in Eq.(\ref{3local}), assuming 
the two-body interactions in Hamiltonian $H^{(2)}$  
are U(1)-invariant, we show that
\begin{align}\label{Eq:Thm}
\Delta_3&=-8 \int_0^T\hspace{-2mm} dt\ \sum_{i<j<k}  a^{(3)}_{ijk}(t)\equiv - 8 \alpha_3 \pmod {2\pi},
\end{align}
where we have defined the phase $\alpha_3=\int_0^T\hspace{-2mm} dt\ \sum  a^{(3)}_{ijk}(t)$.  By measuring the phase $\Delta_3$ and assuming the coefficients $a^{(3)}_{ijk}$ are time-independent, we can measure the net 3-body term $\sum a^{(3)}_{ijk}$, up to an integer multiple of $\pi(4 T)^{-1}$. 
Hence, we can uniquely determine $\sum a^{(3)}_{ijk}$ either by measuring $\Delta_3$ for a sufficiently short time $T$, or, alternatively, by measuring the phase $\Delta_3$ for a few different values of time $T$. Also, if the interactions are translationally or permutationally invariant, this allows us to determine  $a^{(3)}_{ijk}$.

Beyond this example, for a general  U(1)-invariant 
Hamiltonian $H(t)$  in the Supplementay Material we establish the following remarkable identity:
\be\label{gen}
\Delta_3=-\frac{4}{2^n}  \sum_{l\;\text{odd}} (l-1) \int_0^T dt\  \Tr[H(t)C_l]\pmod {2\pi}. 
\ee
Here, the summation is over odd integers and for $l=1,\cdots, n$, the operator $C_l=\sum_{i_1<i_2<\cdots< i_l} Z_{i_1}\cdots   Z_{i_l}$, where  the summation is over $i_1,\cdots, i_l\in\{1,\cdots, n\}$ satisfying $i_1<i_2,\cdots, <i_l$\footnote{For instance, for $n=3$ and $n=4$ qubits operator $C_3=Z_1 Z_2 Z_3$ and $C_3=Z_1 Z_2 Z_3+ Z_1 Z_2 Z_4+Z_1 Z_3 Z_4+Z_2 Z_3 Z_4$, respectively.}. It follows that by  adding the $k$-body interaction  $a^{(k)}_{i_1,\cdots, i_k}(t) Z_{i_1}\cdots Z_{i_k}$ 
acting on $k$ distinct qubits $i_1,\cdots, i_k$ 
to the Hamiltonian $H(t)$, the phase $\Delta_3$ remains unchanged for even $k$, whereas  for odd $k$ it  changes linearly as $$\Delta_3\longrightarrow \Delta_3-4(k-1)\int_0^T dt \ a^{(k)}_{i_1,\cdots, i_k}(t) \pmod {2\pi}.$$ 
This, in particular, implies Eq.(\ref{Eq:Thm}).  It is also worth noting that the phase $\Delta_3$ is additive in time. That is, by repeating the unitary $V$ sequentially $r$ times, which realizes the unitary $V^{r}$, the corresponding phase transforms as  $\Delta_3 \rightarrow r \Delta_3$, which can be useful for amplifying this phase if 3-body interactions are weak.

It is interesting to compare this approach with the original proposed scheme 
\cite{marvian2022restrictions2} based on the notion of $l$-body phases 
\begin{align}\label{def}
\Phi_l\equiv   -\int_0^T\hspace{-2mm} dt\  \  \Tr [H(t) C_l]=\sum_{m=0}^n c_l(m)\theta_m \pmod {2\pi},
\end{align}
where  integer $c_l(m)=\sum_{s=0}^l (-1)^s  {{m}\choose{s}} {{n-m}\choose{l-s}}$ is the eigenvalue of 
$C_l$ in the sector with $m$ excitations.  Note that while each $l$-body phase depends on all  $\theta_m: m=0,\cdots, n$, and therefore its measurement requires probing the unitary $V$ in an exponentially large Hilbert space, to determine $\Delta_3$, one only needs to probe the system in the  $2(n+1)$-dimensional subspace corresponding to the sectors with  $m=0, 1, n-1$, and  $n$ excitations. However, this extra exponential cost of measuring $l$-body phases comes with additional information about the locality of interactions: $\Phi_l$ is only sensitive to $l$-body interactions, i.e., unless  $l=k$, by
adding  the $k$-body interaction   $a^{(k)}_{i_1,\cdots, i_k}(t) Z_{i_1}\cdots Z_{i_k}$ to the Hamiltonian, the $l$-body phase  $\Phi_l$  remains unchanged.
It is also worth noting that in the special case of Hamiltonians in the form of Eq.(\ref{3local}) that only contain up to 3-body interactions,
  comparing Eq.(\ref{gen}) and Eq.(\ref{def}), we find $\Phi_3=2^{n-3}\Delta_3: \pmod {2\pi}$.\\

\color{black} 

\noindent \textbf{The measurement scheme--} Next, we introduce a method for measuring  the phase $\Delta_3$.  Each phase $\theta_m$ is not observable individually,
because it transforms non-trivially under the global phase transformation $V\rightarrow e^{i\phi} V$. To overcome this and find an efficient way of measuring $\Delta_3$,  we rewrite Eq.(\ref{Phase}) as 
\be\label{Eq:2024}
\Delta_3=(\theta_{n-1}-n \theta_n)-(\theta_1-n\theta_{0})+2(\theta_n-\theta_0)
 \pmod{2\pi}.
\ee
Now each of the three phases 
 $ \theta_n-\theta_0$, $\theta_1-n\theta_0$, and $\theta_{n-1}-n \theta_{n}$ remains invariant under the above global phase transformation and
 can be experimentally measured as we demonstrate below. 

\begin{figure}[h]
\centering
\includegraphics[width=0.45\textwidth]{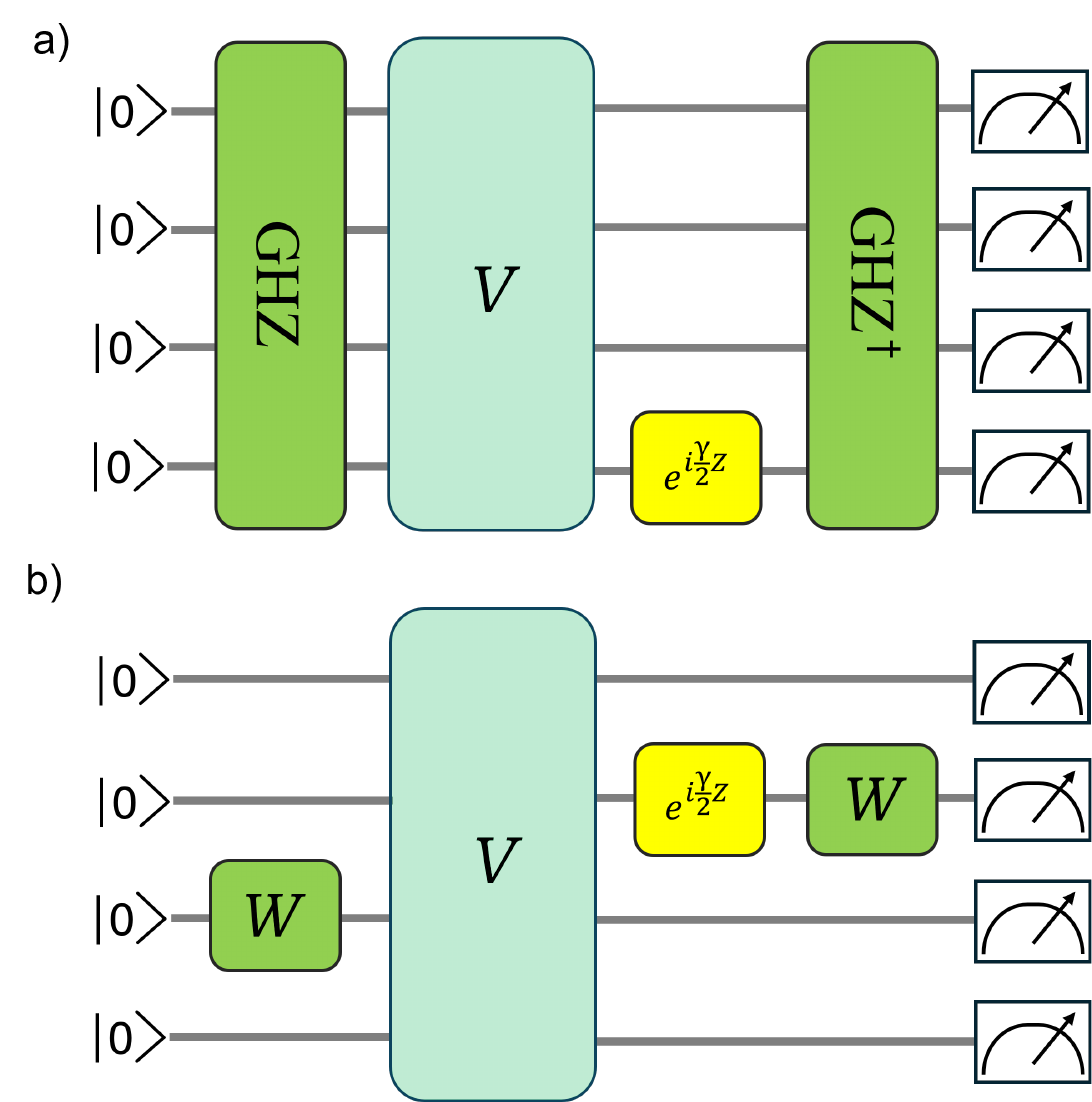}
\caption{\textbf{Circuits for measuring the phase $\Delta_3$:} a) Measurement of $\theta_n-\theta_0$, or the phase between the fully occupied sate $|\textbf{1}\rangle=|1\rangle^{\otimes n}$ and the vacuum $|\textbf{0}\rangle=|0\rangle^{\otimes n}$. The circuit 
first initializes the qubits in state $|\textbf{0}\rangle$ and prepares $|\text{GHZ}\rangle$ (e.g., through a Hadamard gate on the first qubit followed by a sequence of $n-1$ CNOTs controlled by the first qubit). Next, an unknown unitary $V$ is applied, followed by a single-qubit phase shift  $e^{i Z\gamma/2} $ on one of the qubits, then a reversal of the GHZ generation and measurement of the qubits in the computational basis. The probability that all qubits are found in state $|0\rangle$ is given in~Eq.(\ref{prob}). 
b) Measurement of $\theta_1-n \theta_0=\text{arg}(\text{det}(V_1))-n\theta_0$, which can be interpreted as the total phase that single excitation states obtain with respect to the vacuum state. This circuit first initializes qubits in state $|\textbf{0}\rangle$, then applies a Hadamard gate $W$ on qubit $s$, followed by unitary $V$ and a single-qubit phase shift $e^{iZ \gamma/2}$ on qubit $r$, a single-qubit Hadamard $W$ on qubit $r$, and finally measurement of the qubits in the computational basis. The probability of finding all qubits in state $|0\rangle$ is given by Eq.(\ref{prob2}). 
}\label{circuits_GHZ_W}
\end{figure}

First, $\theta_n-\theta_0$ is the relative phase between the fully occupied state $|1\rangle^{\otimes n}$ and the vacuum state $|0\rangle^{\otimes n}$. Thanks to the ability to prepare the GHZ state
\be\nonumber
|\text{GHZ}\rangle=\frac{|0\rangle^{\otimes n}+|1\rangle^{\otimes n}}{\sqrt{2}}\ ,
\ee
which is the equal superposition of these two states, this phase is directly observable via the standard phase sensing methods. Namely, by running the circuit (a) of Fig.~\ref{circuits_GHZ_W}, we measure the probability \be\label{prob}
\big|\langle\text{GHZ}|  e^{i Z_1{\gamma}/{2}}\ V |\text{GHZ}\rangle\big|^2=\frac{1+\cos\big(\theta_n-\theta_0-\gamma\big)}{2}\ ,
\ee
for a few different values of $\gamma$, which uniquely determines  the phase $\theta_n-\theta_0 \pmod{2\pi}$.

Next, consider the phase $\theta_1-n \theta_0=\text{arg}[\text{det}(V_1)]-n\theta_0$, which can be interpreted as the total phases obtained by $n$ single-excitation states relative to the vacuum state $|\textbf{0}\rangle=|0\rangle^{\otimes n}$. 
Of course, standard process tomography allows the determination of $V_1$, up to a global phase. However, in this way, we will not be able to detect the desired relative phase because sensing that phase requires the interference between the 0 and 1 excitation states.
 To achieve this, we use the circuit (b) in  Fig.~\ref{circuits_GHZ_W} to  
measure the probability 
\begin{align}
\big|\langle\textbf{0}| W_r e^{i Z_r\gamma/2} V W_s|\textbf{0}\rangle\big|^2= \frac{1+A^2_{rs}+2 A_{rs}\cos(\alpha_{rs}-\gamma)}{4}\ ,\label{prob2}
\end{align}
where $|\textbf{0}\rangle=|0\rangle^{\otimes n}$ and $W_s$ denotes the Hadamard gate on qubit $s$. Here, $A_{rs}$ $(\alpha_{rs})$ are the absolute value (phase) of the matrix element
\be\label{element}
[\widetilde{V}_1]_{rs}= e^{-i\theta_0} \langle1_r| V|1_s\rangle\equiv A_{rs} e^{i \alpha_{rs}} \ ,
\ee
for $r,s: 1,\cdots, n$,  where $|1_r\rangle= |0\rangle^{\otimes (r-1)} |1\rangle|0\rangle^{\otimes (n-r)}$ 
is the state with a single excitation at qubit $r$. Hence, by repeating this measurement for a few different values of $\gamma$, we can determine  $n^2$ matrix elements $[\widetilde{V}_1]_{rs}$. The determinant of this $n\times n$ matrix determines  the phase $\theta_1-n \theta_0=\arg\text{det}(\widetilde{V}_1)$.  

Finally, the phase $\theta_{n-1}-n \theta_n$, which can be interpreted as 
the total phase obtained by the single-hole states relative to the fully occupied state $|1\rangle^{\otimes n}$, can be measured in a similar fashion using the circuit in Fig.~\ref{circuits_GHZ_W}b, except in this case all qubits should be initialized in state $|1\rangle$. Combining these three phases using Eq.(\ref{Eq:2024}), we find the value of $\Delta_3$, via $\mathcal{O}(n^2)$ different measurement settings.  \\

\begin{figure}[h]
\centering
\includegraphics[width=0.4\textwidth]{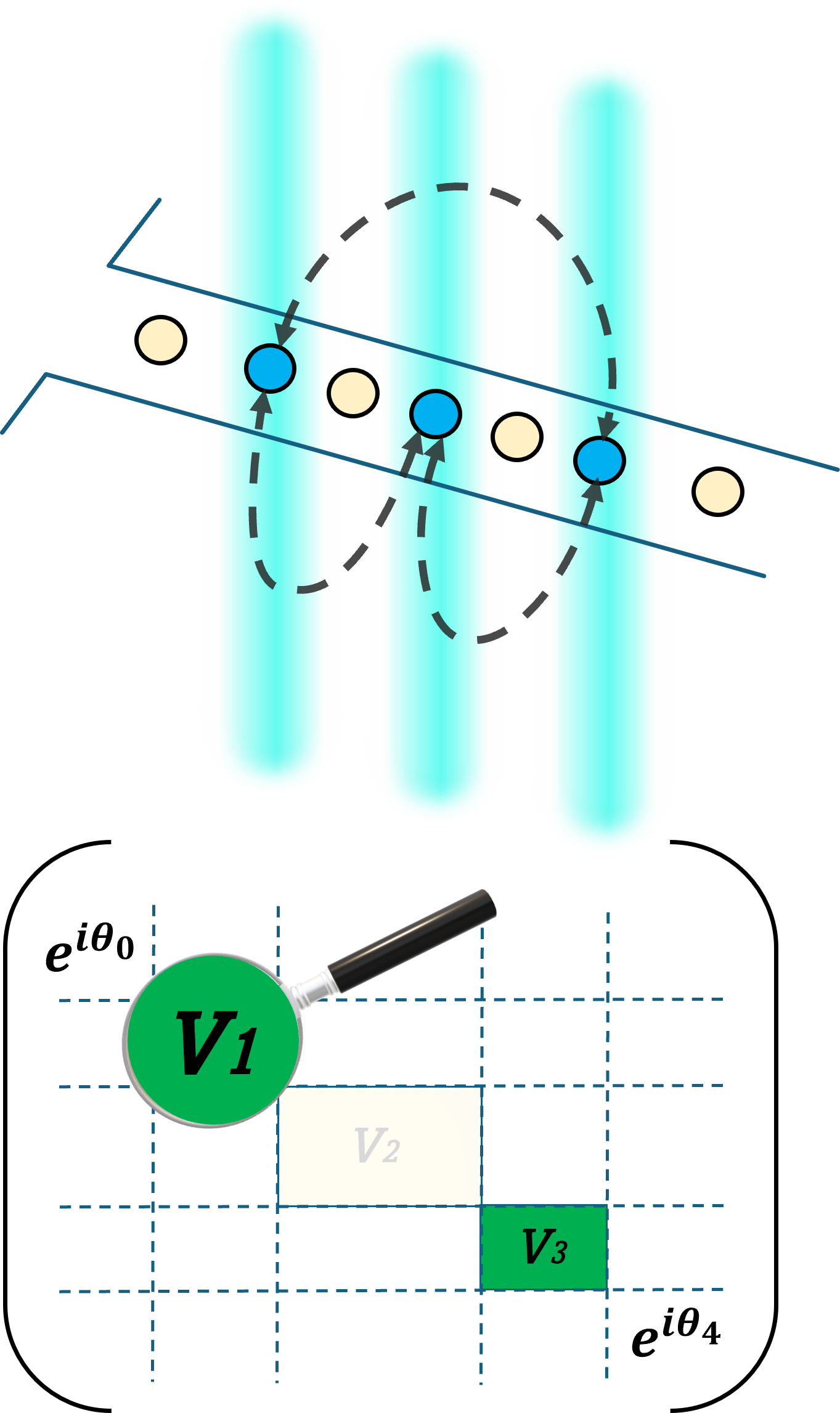}
\caption{\textbf{Schematic presentation of the experiment and the structure of U(1)-invariant unitaries.} We introduce and experimentally demonstrate a novel method for detecting 3-body interactions. 
As shown in the top figure, we implement circuits with 4 and 5 trapped-ion qubits in a linear chain of 7 ions in an RF surface trap. 
 Due to the individual addressing and all-to-all connectivity capabilities, we can execute single-qubit gates on any selected qubits and two-qubit gates on any chosen pair of qubits. 
 The bottom figure represents a schematic depiction  of a U(1)-invariant unitary 
$V$ on $n=4$ qubits. To detect the presence of 3-body interactions, one only needs to 
probe the subspaces with $m=0, 1, n-1, n$ excitations. Specifically for $n=4$ qubits, we do not need to probe the 6-dimensional subspace, with $m=2$ excitations.}\label{experimental_design}
\end{figure}

\noindent\textbf{The experiment--}
 We perform the experiment on 5 qubits within a 7-ion programmable trapped ion quantum computer, with all-to-all connectivity, individual addressing, and efficient state readout of all qubits \cite{H20,egan_BS,Cetina2022} (see Methods for more details). In particular, we use hyperfine qubits of $^{171}$Yb$^{+}$ ions, with single-qubit gate fidelity of 99.6(2)\% and two-qubit gate fidelity ranges from 99.3(1)\% to 98.7(2)\% for the qubit pairs used in this work (the fidelities are \textit{not} corrected for state preparation and measurement [SPAM] errors estimated to be about 0.3\%). 
 
 The ions are arranged in a linear chain, equidistantly spaced, and individually addressed with tightly focused beams passing through a multi-channel AOM, as depicted in the schematic of Fig.~\ref{experimental_design}. The geometry of the individual addressing beams is perpendicular to the plane of the ion chain, enabling the addressing of transverse modes. With such capability, the system can implement universal gates for any qubit or pair of qubits. The native two-qubit gates are M{\o}lmer-S{\o}rensen or Ising operations between arbitrary pairs of qubits \cite{Sorensen:99}.

We employ automated calibration of single- or two-qubit gates and execute the large number of circuits required in the experiment with a periodic evaluation of the system's performance. Namely, we utilize a software-hardware API \cite{wang2024cafqa} alongside an automated calibration procedure within the hardware coding environment. This API optimizes and transpiles gate-level circuits into pulse sequences, then submits these sequences to an integrated hardware control system - ARTIQ (Advanced Real-Time Infrastructure for Quantum Physics)~\cite{bourdeauducq_2016_51303}. ARTIQ manages the coordination of the hardware system to execute the pulse sequences and performs auto-calibrations as needed. By adhering to a software-hardware co-design principle, we enhance circuit execution efficiency and resilience to hardware interruptions, facilitating the handling of a large number of circuits with high performance.\\

\begin{figure}[h]
\centering
\includegraphics[width=0.45\textwidth]{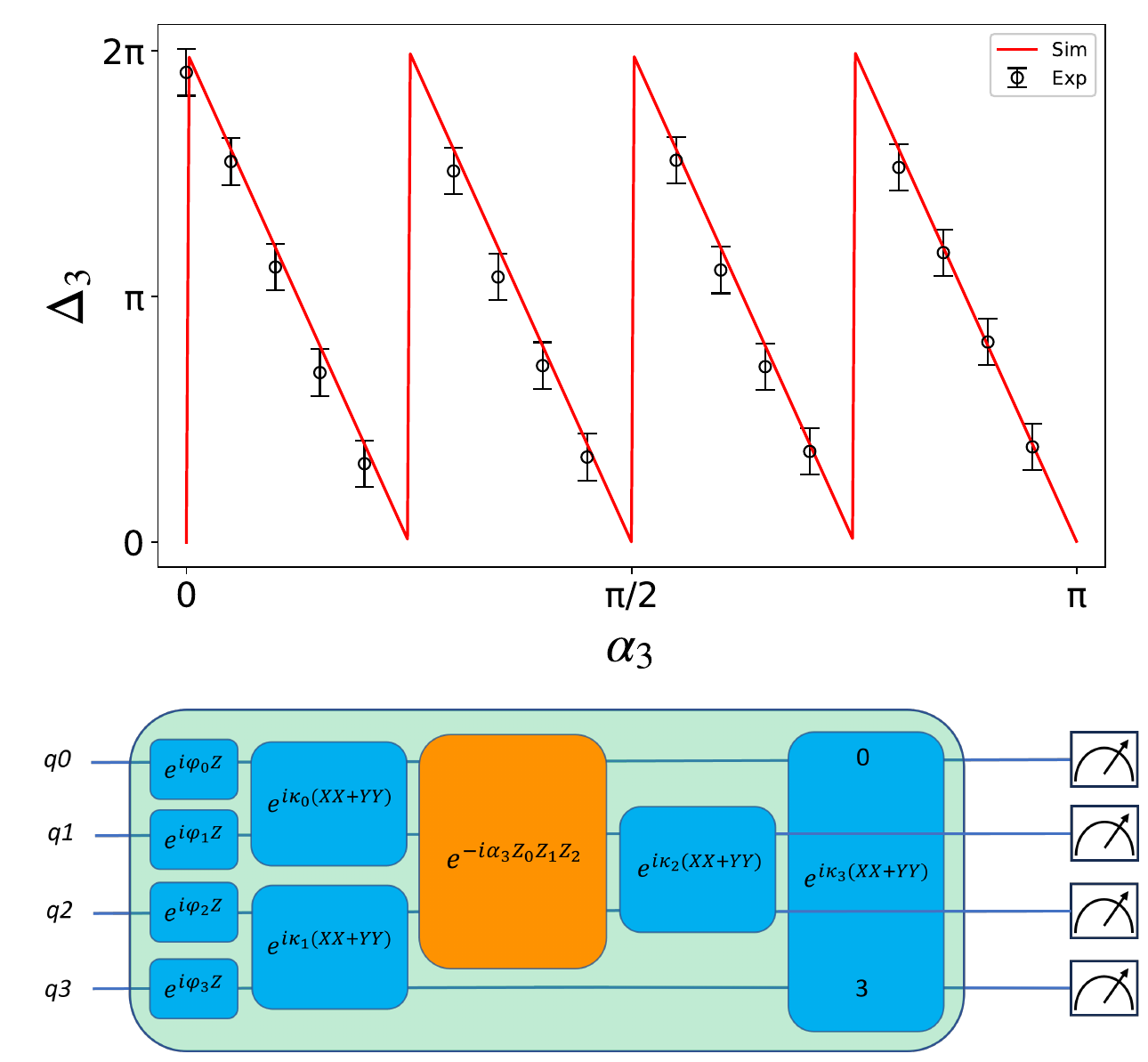}
\caption{\textbf{Measurement of the phase $\Delta_3$}. We perform measurement of $\Delta_3$ for the U(1)-invariant unitary 
realized by the bottom circuit. Except for the unitary $\exp(-\i\alpha_3 Z_0Z_1Z_2)$, the  rest of the gates 
in this circuit
are 1- and 2-qubit U(1)-invariant gates. The 2-qubit gates redistribute the excitations in the system. The plot compares the result of the experiment with 
the theoretical prediction of Eq.(\ref{Eq:Thm}). As predicted, by varying the phase $\alpha_3$ from $0$ to $\pi$, $\Delta_3$ oscillates four times. Angles $\varphi_{0}$, $\varphi_{1}$, $\varphi_{2}$, and $\varphi_{3}$ are $-\pi/5$, $\pi/16$, $3\pi/16$, and $5\pi/14$, whereas $\kappa_{0}$,$\kappa_{1}$,$\kappa_{2}$, and $\kappa_{3}$ are $-\pi/14$,$-\pi/5$,$\pi/6$, and $\pi/9$. The error bars represent two standard deviations ($2\sigma$) determined by the bootstrap method (See Methods for further details). }\label{fig_3body_main}
\end{figure}

\noindent\textbf{Measuring $\Delta_3$--} To demonstrate this method we utilize a setup with four qubits on a trapped-ion quantum simulator (See Fig.~\ref{fig_3body_main}). In addition to one- and two-qubit U(1)-invariant unitaries, the unknown unitary $V$ also contains the three-qubit unitary  $\exp({-i\alpha_3 Z_{0}Z_{1}Z_{2}})$ for $\alpha_3\in(-\pi,\pi]$. 
The two-qubit gates  
exchange excitations between qubits, which means the realized U(1)-invariant unitary is not diagonal in the computational basis.  
The phase $\Delta_3$ for this circuit can be calculated from Eq.(\ref{3local}) to be $
\Delta_3=-8 \int_0^T\hspace{-2mm} dt\ \sum a^{(3)}_{ijk}(t)=-8\alpha_3$ (Note that 
any circuit can be interpreted as a time-dependent Hamiltonian evolution). 

Then, to experimentally  measure $\Delta_3$, we measure the three phases in Eq.(\ref{Eq:2024})
using the circuits in Fig.~\ref{circuits_GHZ_W}. The plot in Fig.~\ref{fig_3body_main} compares the outcome of the experiment with the theoretical prediction in Eq.(\ref{Eq:Thm}). This plot clearly shows a 
distinctive feature of  $\Delta_3$: it oscillates eight times faster than $\alpha_3$. Remarkably, for measuring the phase $\Delta_3$ we do not need to probe the 6D subspace with $m=2$ excitations. We also check that the phase $\Delta_3$ does not change by varying 1-qubit and 2-qubit U(1)-invariant gates~(see Methods).\\





\begin{figure}[h]
\centering
\includegraphics[width=0.4\textwidth]{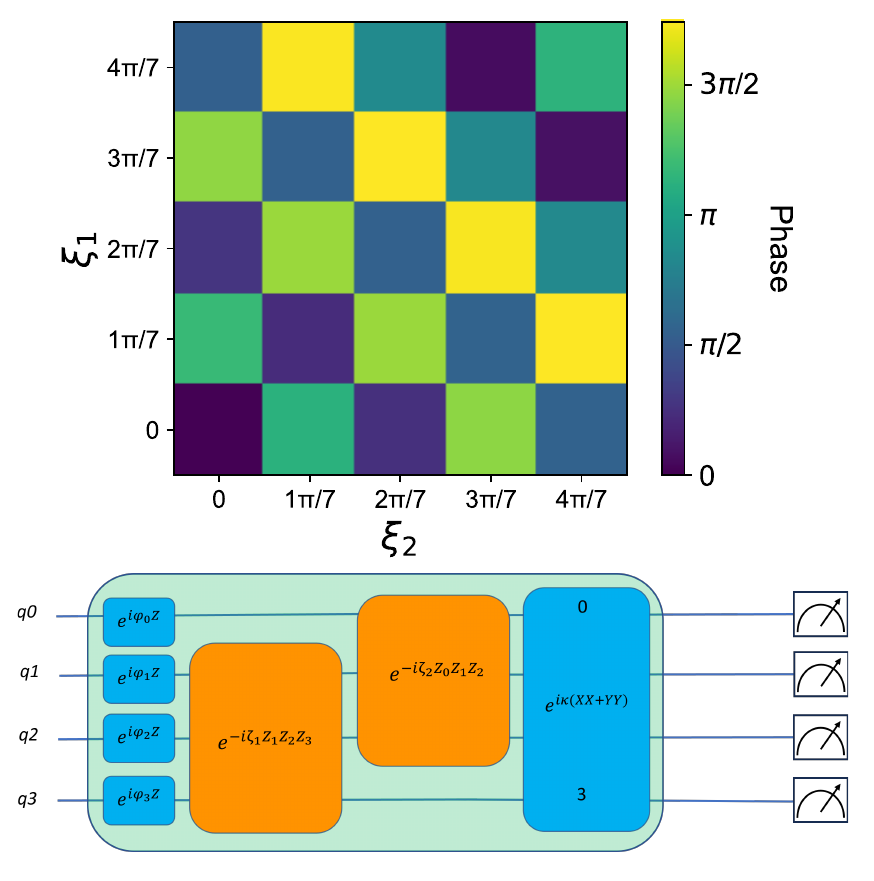}
\caption{\textbf{Experimental verification of the additivity of phase $\Delta_3$.} The circuit contains two 3-qubit gates $\exp(-\i \xi_1 Z_1Z_2Z_3)$ and  $\exp(-\i \xi_2 Z_0Z_1Z_2)$. As predicted by Eq.(\ref{Eq:Thm}), the phase $\Delta_{3}$ is $-8 (\xi_1+\xi_2): \text{mod}\ 2\pi$, which is consistent with the off-diagonal pattern in this plot. In this experiment, we set single-body z-rotation angles $\varphi_{0}$, $\varphi_{1}$, $\varphi_{2}$, and $\varphi_{3}$ to be $\pi/5$, $\pi/16$, $\pi/3$, and $\pi/12$, whereas $\theta$ is $-3\pi/10$ for two-qubit unitary. The error for each measured $\Delta_{3}$ is $\pm 0.2$ determined by the bootstrap method (See Methods for further details).} \label{fig_3_and_3}
\end{figure}

\noindent\textbf{Additivity of $\Delta_3$--} 
Another crucial feature of the phase $\Delta_3$ is additivity: according to Eq.(\ref{Eq:Thm}) all 3-body interactions have equal contributions in the phase   
$\Delta_3$. To experimentally verify this we consider the unitary $V$ in the following form {changed}
\begin{equation}
V=S_2    e^{-i \zeta_2 Z_1Z_2Z_3}e^{-i \zeta_1 Z_0Z_1Z_2} S_1\ , \label{eq:V_add}   
\end{equation}
on a system with $n=4$ qubits, where $S_1$, and $S_2$ are random 2-local U(1)-invariant unitaries, namely $\exp(i \phi Z_j)$, and $\exp(i \phi [X_{j}X_k+Y_jY_k])$ for random values of $\phi$. As it is shown in Fig.~\ref{fig_3_and_3}, by varying the 3-body interaction strength $\zeta_1$ and $\zeta_2$, we experimentally verify the additivity of the phase $\Delta_3$. Specifically, we demonstrate that  $\Delta_3=-8(\zeta_1+\zeta_2)\pmod {2\pi} $, which follows from Eq.(\ref{Eq:Thm}).\\

\noindent\textbf{Measuring $l$-body phases--}  In the last experimental setup, we measure the $l$-body phases in Eq.(\ref{def}), which unlike $\Delta_3 $ allows us to distinguish $k$-body interactions, with different $k$. However, this requires performing full tomography by probing all sectors of unitary $V$.

\begin{figure}[h]
\centering
\includegraphics[width=0.4\textwidth]{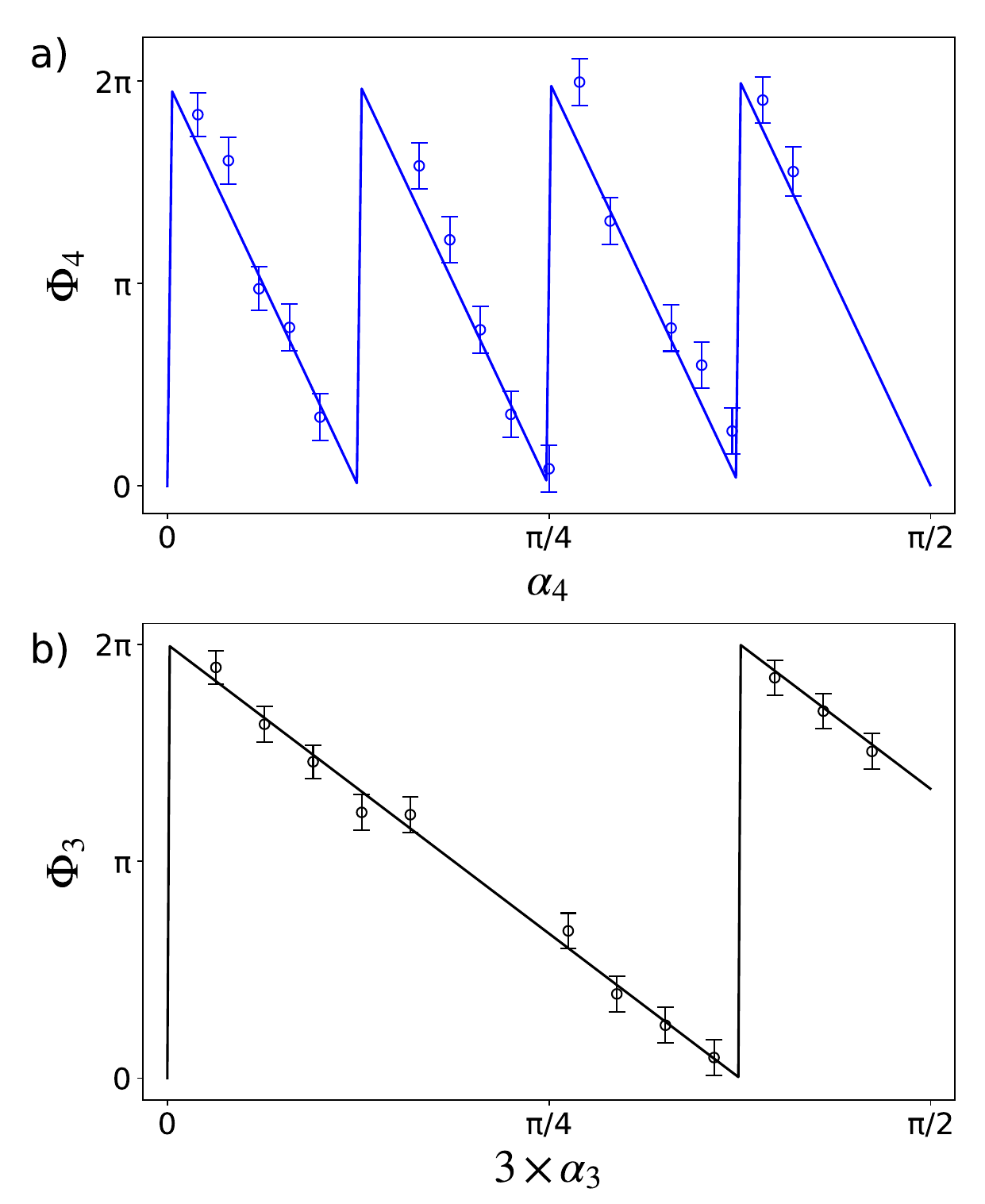}
\caption{\textbf{Measurement of $3$-body and $4$-body phases $\Phi_3$ and $\Phi_4$.} In this experiment we measure $\Phi_3$ and $\Phi_4$ defined in Eq.(\ref{def}), on $n=4$ qubits for 
the unitary realized by the circuit in Fig.~\ref{fig_phi_circuits}. To measure these quantities we have performed the full process tomography for this unitary. The circuit contains 4-qubit unitary $\exp(-\i\alpha_4 Z_0Z_1Z_2Z_3)$ and the 3-qubit unitary $\exp(-\i\alpha_3 Z_1Z_2Z_3Z_4)$, where for convenience in this circuit we have selected them such that $\alpha_4=3\alpha_3$ (i.e., we assume the 3-body interaction is three times weaker than the 4-body interaction). 
Eq.(\ref{def}) implies that $\Phi_3=2^4 \alpha_3$, and $\Phi_4=2^4 \alpha_4$, which correspond to the solid black and blue lines in the above plots. Dots represent experimentally measured phases. The error bars on the figure represent two standard deviations ($2\sigma$) determined by the bootstrap method.}
\label{fig_phi}
\end{figure}

In this experiment, 
we consider a system with $n=4$ qubits evolving under
the unitary  
\begin{equation}\nonumber
V= e^{-\i \alpha_{4} Z_0Z_1Z_2Z_3}  e^{-\i {\alpha_3} Z_1Z_2Z_3} S_2  S_1\ ,    
\end{equation}
where $S_1$ and $S_2$ are random one- and two-qubit  U(1)-invariant unitaries. To simplify the experiment and presentation, rather than varying $\alpha_3$ and $\alpha_4$ independently, 
we choose a particular ratio, namely $\alpha_3=\alpha_4/3$.
This can be interpreted as the unitary corresponding to the Hamiltonian $Z_0Z_1Z_2Z_3+\frac{1}{3}Z_1Z_2Z_3$ plus one- and two-qubit U(1)-invariant terms. 
  Fig.~\ref{fig_phi} presents the corresponding 3-body and 4-body phases, $\Phi_{3}$ and $\Phi_{4}$. As expected  from Eq.(\ref{def}),  both $\Phi_4$ and $\Phi_3$  oscillate $2^n=16$ times faster than $\alpha_4$ and $\alpha_3$, respectively.
 
To perform the process tomography on $V$, we use a technique that takes advantage of the U(1) symmetry of this unitary. 
 On a system with $n$ qubits, this symmetry implies that, instead of $4^n$ real parameters, to specify the unitary we need to determine $\sum_{m=0}^n {{n}\choose{m}}^2={{2n}\choose{n}}\approx 4^n/\sqrt{\pi {n}}$ \cite{bai2023synthesis} (For the present experiment this means $ {{8}\choose{4}}=70$ real numbers instead of $4^4=256$).  Therefore, to take advantage of the symmetry constraints, rather than the standard 
tomography protocols which ignore this structure, we utilize a different technique. Namely, we use variants of the circuit (b) in  Fig.~\ref{circuits_GHZ_W}, with more Hadamard gates at the input and output, which allows the creation of more excitations. \\


\section*{Discussion}

We introduced a new method for observing the symmetry-protected signature of 3-body interactions and performed the first-ever experiment demonstrating its measurement, enabling the direct detection of 3-body interactions.  Our experiment was conducted on a trapped ion QC with an average 2-qubit gate error of 1$\%$. The possibility of such direct detection of 3-body interactions was recently revealed in \cite{marvian2022restrictions2}. While the original method proposed in \cite{marvian2022restrictions2} requires full process tomography, the new scheme introduced in this paper only probes a subspace that scales linearly with the number of qubits. To achieve this, we take advantage of GHZ states, which allow us to directly measure the relative phase between the vacuum and fully occupied states. Without  GHZ states, the presence of 3-body interactions can still be detected, although the efficiency will generally be lower. For instance, in Methods, we introduce a different scheme that does not use GHZ states; however, it requires probing the system in the subspace with 1, 2, and 3 excitations, whose total dimension scales as $n^3$. This alternative scheme can be useful when preparing GHZ states is challenging, e.g., when the excitations are massive.\\

\newpage

\section*{Methods} 

\section{Path-independent observable phases protected by symmetry}
The phase $\Delta_3$ introduced in Eq. (\ref{Phase}) is an example of a general formalism introduced in Ref. \cite{marvian2022restrictions2} for constructing path-independent phases under symmetric Hamiltonians. In the following, first, we review this formalism and then study the properties of phase $\Delta_3$. Finally, in Sec. \ref{Sec:alt}, we introduce another family of such phases, which provides an alternative approach for detecting 3-body interactions.

\subsection{Review: The formalism of the symmetry-protected path-independent phases}

Consider a system with $n$ qubits with the total Hilbert space $(\mathbb{C}^2)^{\otimes n}$. 
 Let $\Pi_m$ be the projector to the eigen-subspace of $\sum_{j=0}^n Z_j$ with eigenvalue $n-2m$, i.e., the subspace with $m$ excitations. Suppose this system evolves under  U(1)-invariant Hamiltonian $H(t): 0\le t\le T$, which   realizes the unitary $V=\mathcal{T}\{\exp(-\i\int_0^T dt H(t))\}$. 
Then, U(1) symmetry  implies that $H(t)$ and $V$ decompose
as $\bigoplus_{m} H_m(t)$ and
$\bigoplus_{m} V_{m}$, where $H_m$ and $V_m$ are are components of $H$ and $V$
 in the subspace with $m$ excitations, respectively. Furthermore, 
 \bes
\begin{align}
V_m&=\mathcal{T}\left\{\exp\left[-\i \int_{0}^T H_m(t) dt\right] \right\}\\ &=\lim_{L\rightarrow \infty}\prod_{j=1}^L \exp({-\frac{\i T}{L} H_m(\frac{T j}{L}) })\ .
\end{align} 
\ees
Next, recall that for any pair of  operators $A_1$ and $A_2$, $\text{det}(e^{A_1}e^{A_2})=e^{\Tr(A_1)+\Tr(A_2)}$, which implies   
\be
\text{det}(V_m)=\exp\left({-\i\int^T_0  \Tr(H(t)\Pi_m) dt}\right)\ .
\ee
Defining $\theta_m=\text{arg}[\text{det}(V_m)]$, i.e., the phase of $\text{det}(V_m)$, we conclude  that  for any set of integers $\{c(m)\}$, 
\be\label{ary2}
\Phi\equiv\sum_{m}   c(m)\ \theta_m \hspace{-1mm} =-\hspace{-1mm}\int^T_0 \hspace{-1mm}dt\ \Tr[H(t)C]\pmod {2\pi},  
\ee
where   $C=\sum_{m} c(m)\Pi_m$. Note that $\theta_m$ is defined only mod $2\pi$. Therefore, the phase $\Phi$  is well-defined only when coefficients $c(m)$ are integers. The above argument, in particular, means that if Hamiltonians $H_1(t)$ and $H_2(t)$ generate the same unitary $V$, then 
\be
\int_0^T dt\  \Tr[H_1(t)C]=\int_0^T dt\  \Tr[H_2(t)C]\pmod {2\pi}.
\ee  
If  operator $C$ is traceless then $\Phi$ remains invariant under a global phase transformation $V\hspace{-1mm}\rightarrow\hspace{0mm} e^{\i\alpha} V$ for any $\alpha\in[0,2\pi)$. This can be seen, e.g., using the 
 fact that if Hamiltonian $H(t): 0\le t\le T$ realizes $V$, then Hamiltonian  $H(t)-\frac{\alpha}{T} I: 0\le t\le T$ realizes $e^{i\alpha} V$. For traceless operator $C$, $\Tr(H(t) C)$ remains invariant under this transformation. Then,  the second equality in Eq.(\ref{ary2}) implies that the phase $\Phi$ remains invariant.

In summary, for any traceless operator $C=\sum_m c(m)  \Pi_m$ with integer eigenvalues $\{c(m)\}$, the phase $\Phi$ defined in Eq.(\ref{ary2}) is an  observable quantity.  As we see next, by measuring this phase we can obtain information about the locality of the Hamiltonian $H(t)$.

As an example, recall the definition of operators $\{C_l\}$  in \cite{marvian2022restrictions2}: Operator $C_0=I^{\otimes n}$ is defined to be the identity operator on $n$ qubits, and for $l=1,\cdots, n$ 
\be\label{C3}
C_l=\sum_{i_1<i_2<\cdots< i_l} Z_{i_1}\cdots   Z_{i_l} \ ,
\ee
where the summation is over $i_1,\cdots, i_l\in\{1,\cdots, n\}$ that satisfy $i_1<i_2,\cdots, <i_l$.   Operators $C_l$ decomposes as  
\be\label{Fourier}
C_l=\sum_{m=0}^n c_l(m)\Pi_m\ ,
\ee
where eigenvalues 
\be\label{eigen}
 c_l(m)=\sum_{s=0}^l (-1)^s  {{m}\choose{s}} {{n-m}\choose{l-s}}\ \ \ \ \ \ \ :\ m=0,\cdots, n\ ,
\ee
are all integers \cite{marvian2022restrictions2}.  Ref.\cite{marvian2022restrictions2} defines the $l$-body phase $\Phi_l$ associated to the time evolution generated by the U(1)-invariant Hamiltonian $H(t)$, as  
\begin{align}
\Phi_l\equiv   -\int_0^T\hspace{-2mm} dt\  \  \Tr[H(t) C_l]=\sum_{m=0}^n c_l(m)\theta_m \pmod {2\pi},
\end{align}
where $\theta_m=\text{arg}(\det(V_m))$.  As explained before,  by  adding the $k$-body interaction  $a^{(k)}_{i_1,\cdots, i_k}(t) Z_{i_1}\cdots Z_{i_k}$ 
acting on $k$ distinct qubits $i_1,\cdots, i_k$ 
 to the Hamiltonian $H(t)$, $\Phi_l$ remains unchanged for all $l$, except $l=k$. 
 Therefore, by measuring $\Phi_l$ we can detect $l$-body interactions. See also \cite{marvian2024rotationally}  for the extension of this formalism to the case of SU(2) symmetry.


\subsection{Localization in frequency and spatial domains}\label{Sec:local}

The above property of operators $\{C_l\}$ and the corresponding phases $\{\Phi_l\}$ means that they provide sharp information about the locality of interactions. However,  because for each integer $l$ the corresponding integers $\{c_l(m)\}$ are typically non-zero, to measure $l$-body phase $\Phi_l$, we need to probe the entire Hilbert space, i.e., sectors with arbitrary excitation numbers $m=0,\cdots, n$.

The operators $\{C_l: l=0,\cdots, n\}$ and projectors $\{\Pi_m: m=0,\cdots, n\}$ are two different orthogonal bases, with respect to the Hilbert-Schmidt inner product, for the same operator space. That is, 
$\text{span}\{C_l\}_l=\text{span}\{\Pi_m\}_m\ ,$ and $$\Tr(C_lC_{l'})=\delta_{l,l'}\times \Tr(C_l^2)=\delta_{l,l'} \times 2^n {{n}\choose{l}}\ $$
and $$\Tr(\Pi_m\Pi_{m'})=\delta_{m,m'}\times \Tr(\Pi_m)=\delta_{m,m'}  \times {{n}\choose{m}}\ ,$$
for all $l,l',m,m'=0,\cdots, n$. Hence, Eq.(\ref{Fourier}) describes a change of basis from one orthogonal basis to another. Furthermore, while the basis 
 $\{C_l\}$ defines a sharp notion of locality, the basis  $\{\Pi_m\}$ defines a sharp notion of the number of excitations (charge, or Hamming weight) in the system.  The fact that integers $\{c_l(m)\}$ are typically non-zero, means that each element of the $\{C_l\}$ basis has support on many elements of the basis $\{\Pi_m\}$.  
  In this sense, the transformation defined by Eq.(\ref{Fourier}) 
is analogous to the Fourier transform, and the relation between $\{\Pi_m\}$  and $\{C_l\}$ bases is analogous to the Fourier uncertainty relation.

\begin{figure}[h]
\centering
\includegraphics[width=0.475\textwidth]{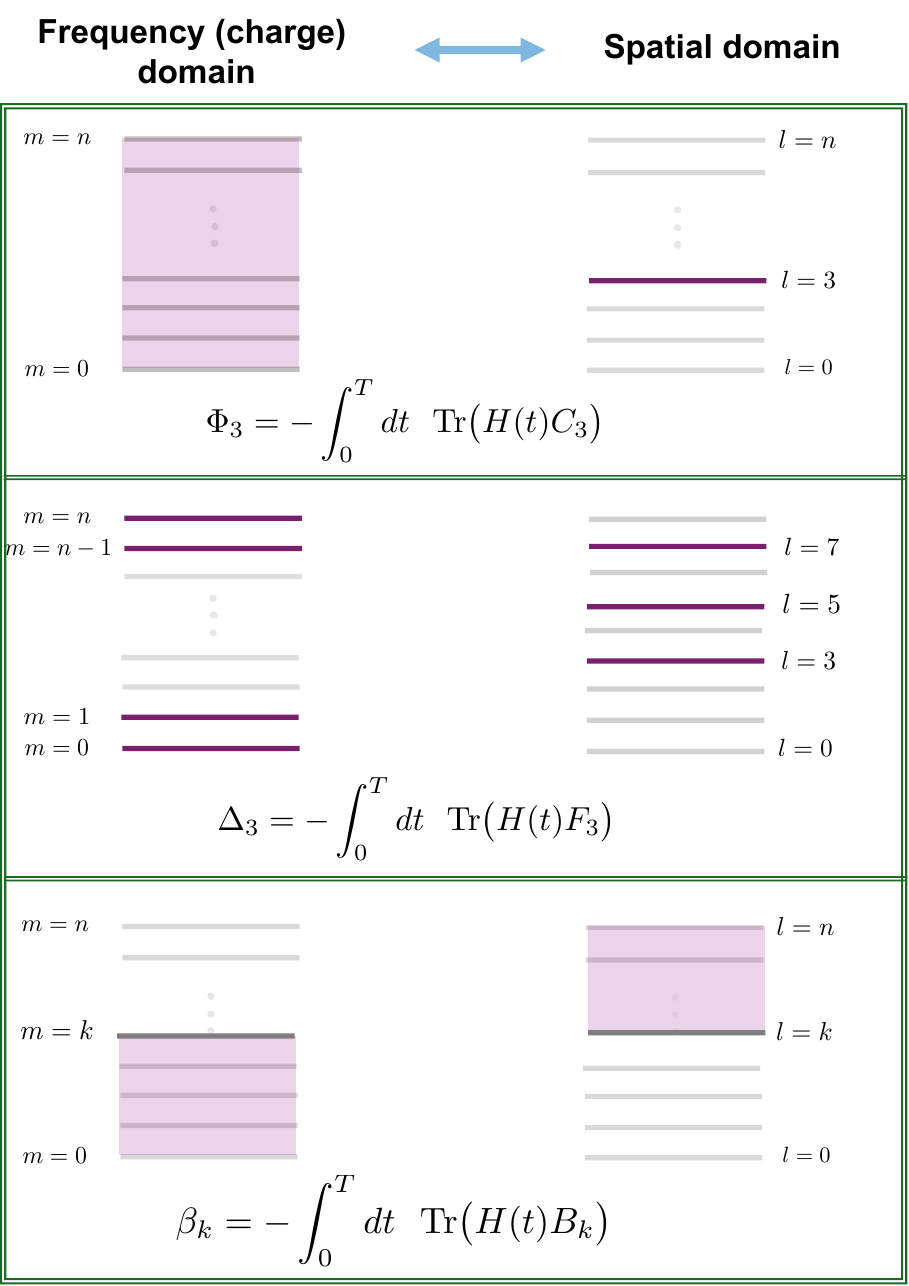}
\caption{\textbf{Frequency versus spatial domains}--The representation of operators $C_3$ defined in Eq.(\ref{C3}), $F_3$ defined in Eq.(\ref{F3}), and $\{B_k\}$ defined in 
Eq.(\ref{Bk}), in the $\{\Pi_m\}$ basis (left) and  $\{C_l\}$ basis (right). See the discussion in Sec.\ref{Sec:local}.
}\label{domains}
\end{figure}

As we explain next, from this point of view the phase $\Delta_3$, and the phases $\beta_k$ defined in Sec.\ref{Sec:alt}, correspond to operators that are partially localized in both spatial and frequency (charge) domains (See Fig.~\ref{domains}).

\subsection{$\Delta_3$ as a path-independent phase}  

Suppose in Eq.(\ref{ary2}), we choose $C$ to be the operator
\be\label{F3}
F_3=(n-2)(\Pi_0-\Pi_n)- (\Pi_1-\Pi_{n-1})\ ,
\ee
which has integer eigenvalues and is traceless. Then, applying Eq.(\ref{ary2})  we find that 
\begin{align}\label{defDel}
\Delta_3&= -\int_0^T\hspace{-2mm} dt\  \  \Tr[H(t) F_3]\\ 
&=(n-2)\times (\theta_0-\theta_n)- (\theta_1-\theta_{n-1})\pmod {2\pi}.\nonumber
\end{align}

To understand properties of the phase $\Delta_3$, in Appendix \ref{App:A} we find the decomposition of operator  $F_3$ in terms of 
$\{C_l\}$ operators. Namely, we show that 
\be
F_3=-\frac{4}{2^n}  \sum_{l: \text{odd}} (l-1) C_l \ , 
\ee
where the summation is over odd integers in the interval  $l=1,\cdots, n$ (See the middle part of Fig.~ \ref{domains} for a schematic representation of this equation).  Putting this into Eq.(\ref{defDel}) immediately implies Eq.(\ref{gen}). Furthermore, in the special case of Hamiltonians that can be decomposed as a sum of 3-body interactions, such as Hamiltonian $H^{(3)}(t)$ in Eq.(\ref{3local}), we have $\Tr(C_l H_{2\text{-loc}})=0$ for $l>2$. Moreover, for distinct $i, j, k$, 
$$\Tr(C_l Z_iZ_jZ_k)=2^n \delta_{3,l}\ .$$  
This implies that for the time evolution under Hamiltonian in Eq.(\ref{3local}), we have
 \begin{align}\nonumber
\Delta_3=-8 \int_0^T dt\ \sum_{i<j<k}  a^{(3)}_{ijk}(t)\pmod {2\pi},
 \end{align}
which proves  Eq.(\ref{Eq:Thm}). 
In this case 
\be\label{gh}
\Phi_3=2^{n-3}\times \Delta_3 \pmod {2\pi},
\ee
which can be obtained by comparing Eq.(\ref{gen}) with Eq.(\ref{def}).  



\subsection{An alternative approach for detecting 3-body interactions}\label{Sec:alt}

As we saw before, to measure the phase $\Delta_3$ one needs to probe 
the subspaces with $m=0, 1, n-1, n$ excitations. We saw that this can be achieved efficiently, provided that one can prepare GHZ states that contain superposition of states with $m=0$ and $m=n$ excitations. However, for some applications, preparing such a superposition might be unfeasible (e.g., when the excitations are massive).

For such applications, here we introduce another
family of path-independent phases, which can be measured by probing the system in the sectors with $m=0, 1, 2, 3$ excitations.  Indeed, we introduce a more general formalism that allows us to detect $k$-body interactions by probing the sectors with $m=0,\cdots, k$ excitations.   

For $k\le n$, define the phase 
\begin{align}\label{defbeta}
\beta_k&\equiv \sum_{m=0}^k (-1)^m {{n-m}\choose{k-m}} \ \theta_m\\ &=-\int_0^Tdt\  \Tr(B_k H(t))\pmod {2\pi} \nonumber \ ,
\end{align}
where operator
\be\label{Bk}
B_k=\sum_{m=0}^k (-1)^m {{n-m}\choose{k-m}} \ \Pi_m\ .
\ee
This definition, in particular, means that operators $\{B_k\}$ are linearly independent, and form a basis for the operator space
\be
\text{span}\{\Pi_m\}=\text{span}\{C_l\}=\text{span}\{B_k\}\ .
\ee
Note that $\beta_k$ only depends on $\theta_0,\cdots ,\theta_k$.  
For instance, for $k=3$, we find
\begin{align}
\beta_3={{n}\choose{3}} \theta_0-{{n-1}\choose{2}} \theta_1+(n-2) \theta_2-\theta_3 \pmod {2\pi} \nonumber\ .
\end{align}
For the special case of $n=3$ qubits, we have 
\be
n=3:\ \  \  \  \   B_3=F_3=C_3=Z^{\otimes 3}\ ,
\ee
which implies   $\beta_3=\Delta_3=\Phi_3$.

The crucial property of operators $\{B_k\}$,  which is proven in the Supplementary Material, is the following:  For all $k,l=0,\cdots, n$ we have 
\bes\label{Bk0}
\begin{align}
\Tr(B_k C_l)&=0\ \ \ \ \ \  : l<k\\
\Tr(B_k C_k)&=2^{k-n} \times \Tr(C_k^2)=2^{k}  {{n}\choose{k}} \ .
\end{align}
\ees
The bottom part of Fig.~ \ref{domains} presents a schematic representation of Eq.(\ref{Bk0}) and Eq.(\ref{Bk}).  In particular, except $k=0$, which corresponds to $B_0=|0\rangle\langle 0|^{\otimes n}$,  the rest of operators $\{B_k\}$ are traceless, which means $\beta_k: k=1,\cdots, n$ are experimentally observable. Note that according to Eq.(\ref{defbeta}),  to measure $\beta_k$, we only need to probe sectors with $m=0, \cdots, k$ excitations.

Furthermore, applying this equation, in  Appendix \ref{App:B}, we show that 
\begin{lemma}\label{lem2}
Suppose operator $H$ can be written as a sum of $k$-local terms. Then,  
\be
\Tr(H B_k)=\frac{2^{k} \times \Tr(H C_k)}{2^n}
\ ,
\ee
and $\Tr(H B_r)=0$ for $r>k$\ .
\end{lemma}
This lemma implies that for U(1)-invariant  Hamiltonian $H$ in the form of Eq.(\ref{3local}), 
\begin{align}
\beta_3=\Delta_3=-8 \int_0^T dt\ \sum_{i<j<k}  a^{(3)}_{ijk}(t)\pmod {2\pi}.
 \end{align}
More generally, consider the Hamiltonian 
\be
H(t)=H_{(k-1)\text{-loc}}(t)+\sum_{j_1<\cdots <j_{k}} a^{(k)}_{j_1\cdots j_k}(t)\ Z_{j_1}Z_{j_2}\cdots Z_{j_k}\ ,
\ee
where $H_{(k-1)\text{-loc}}(t)$ is an unknown U(1)-invariant Hamiltonians that does not contain $k$-body  interactions, i.e., can be decomposed as the sum of $(k-1)$-local U(1)-invariant terms. For this Hamiltonian, we obtain 
\be
\beta_k=-2^k \int_0^T dt\ \sum_{j_1<\cdots <j_{k}}  a^{(k)}_{j_1\cdots j_k}(t) \pmod {2\pi}.
\ee

\section*{Further details and error analysis for the experiment}

\noindent\textbf{Trapped Ion Quantum Computer:} We decompose the transverse modes of 
seven $^{171}$Yb$^{+}$ ions and estimate the effect of noise and drifts on the gates \cite{PhysRevA.97.062325}. This analysis is followed by constructing segmented optical pulses tailored for the entangling gates to ensure their robustness. Gate duration and pulse shaping are adjusted for individual ion pairs to minimize coupling to motion at the end of the gate pulse sequence.

The two-qubit gates fidelity measurement technique is outlined in \cite{egan_BS}. A single-qubit gate is a composite SK1 pulse \cite{PhysRevA.70.052318}. For SPAM measurement, we prepare qubits in the $\left|0\right>$ (dark) and $\left|1\right>$ (bright) states, then perform a Poisson fit to the histograms corresponding to these two states. The resulting SPAM is 0.27(4)\%.

\noindent\textbf{Circuit Design and Optimization:}
For all circuits, commuting single-qubit rotations and two-qubit XX($\pi$/4) gates are merged to minimize the circuit size. However, while the unitary operator $V$ is still subject to optimization, its content is not merged with the external circuit elements and is treated as a `black box'. The 3-body and 4-body  unitaries $\exp(-\i \alpha_3 Z_0Z_1Z_2)$ and $\exp(-\i \alpha_4 Z_0Z_1Z_2Z_3)$ are realized using standard methods for implementing Pauli Hamiltonians, via a sequence of CNOTs \cite{Nielsen:01}.

\noindent\textbf{Post-Processing and Statistical Error:}
Data presented in the main text are the result of several post-processing steps. First, we extract the population from the raw histograms for different values of $\gamma$. Then we perform fitting to extract observable phases. The next step is a construction of the total phases corresponding to 3- or 4-body interaction following Eq.(\ref{Phase}) or Eq.(\ref{def}), respectively.

\begin{figure}{\includegraphics[width=0.4\textwidth]
{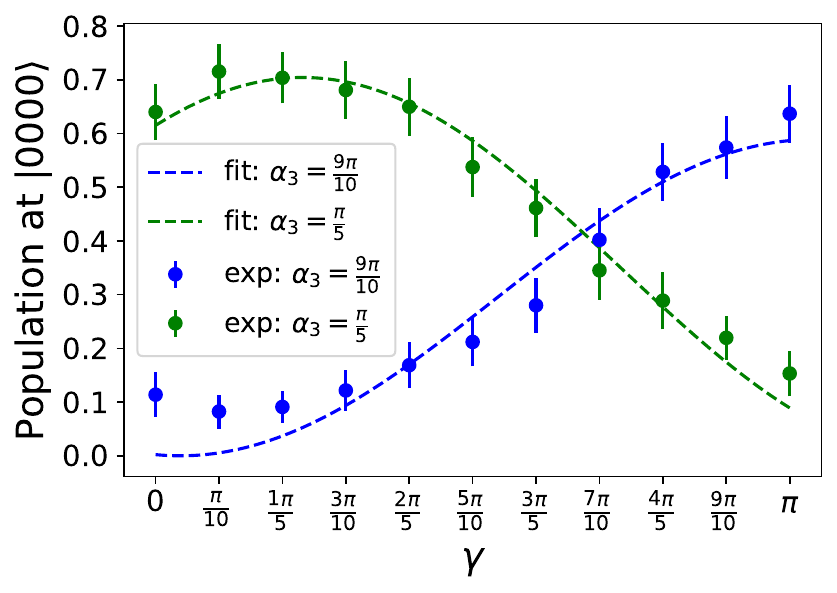}}\centering\caption{
\textbf{Phase estimation using 4-qubit GHZ state}
Using the circuit in Fig.~\ref{fig_3body_main}, we measure the phase $\theta_{4}-\theta_{0}$ via Eq.~(\ref{prob}). We present two setups corresponding to three-qubit unitaries  $\exp({-\i\alpha_3 Z_{0}Z_{1}Z_{2}})$  with $\alpha_{3}=\pi/5$ and $\alpha_{3}=9\pi/10$. For each value of interaction strength $\alpha_{3}$, we vary $\gamma$ in Eq.(\ref{prob}). By fitting the resulting curves with sinusoidal functions, we extract the phase offsets corresponding to the phase $\theta_{4}-\theta_{0}$. See the caption of Fig.~\ref{fig_3body_main} for further details on the realized 
unitary $V$ in this experiment. The error bars represent two standard deviations ($2\sigma$) determined by the bootstrapping method discussed below.} 
\label{post_process_example}
\end{figure}

Fig.~\ref{post_process_example} demonstrates the outcomes of the phase estimation experiment for measuring $\theta_4-\theta_0$, with 4-qubit GHZ state. Specifically, we measure the population at $|0000\rangle$ for two distinct interaction strengths of the 3-body interaction, $\alpha_{3}=\pi/5$ and $\alpha_{3}=9\pi/10$. For each interaction strength, we vary $\gamma$ (see Eq.(\ref{prob})), then we perform fitting with a sinusoidal function to obtain the contrasts and ensure the correct measurement of the phase offset. According to Eq.(\ref{prob}), the measured phase offset determines the phase $\theta_4-\theta_1$ (A similar technique is used to determine the phase $\alpha_{rs}$ via Eq.(\ref{prob2})). In this particular example, we measure $\theta_{4}-\theta_{0}$ referenced in~Eq.(\ref{prob}). We use a similar procedure followed by Eq.(\ref{prob2}) to measure the rest of the observable phases.

We use a statistical bootstrapping method~\cite{Bootstrap} to estimate uncertainties by 1000 expectation values with each value evaluated based on a 150-shot histogram randomly resampled from the original 300-shot readout. The error bars indicate 2$\sigma$ from the bootstrap distribution.

\noindent\textbf{Noise Model:} We examine hardware error sources by directly measuring errors and simulating hardware output using the Qiskit platform~\cite{qiskit2024}. We describe a two-qubit gate error in the $z$- and $x$-bases. For all ion pairs used in this work, the errors measured in the $z$-basis are in the range from 0.28(5)$\%$ to 0.85(4)$\%$, whereas errors in the $x$-basis are in the range from 0.01(1)$\%$ to 0.04(1)$\%$. The dominant noise in this model is incoherent (Pauli) error.

\begin{figure}[h]
\centering
\includegraphics[width=0.4\textwidth]{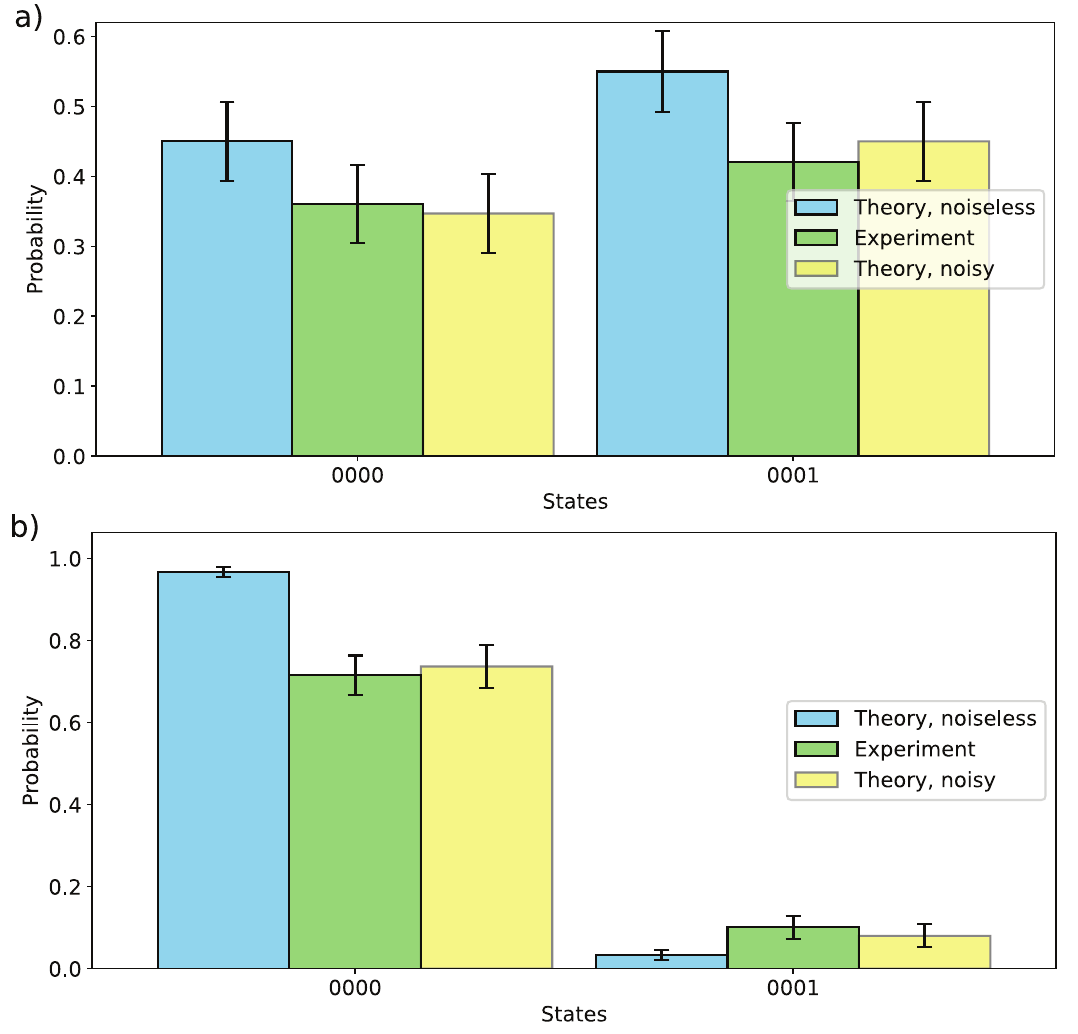}
\caption{\textbf{Error model evaluation.} Outcome of GHZ experiment/simulation on 4 qubits with (a) $\alpha=0.02\pi$ and $\gamma=\pi$, and (b) $\alpha=0.165\pi$ and $\gamma=2\pi/5$. For clarity, only $|0000\rangle$ and $|0001\rangle$ states are presented. All histograms depict output data with 300 shots. The `Theory, noisy' histogram is compared to the `Experiment' histogram to assess the noise model. The error bars on the figure represent two standard deviations ($2\sigma$) determined by the bootstrapping method discussed below.}\label{noise}
\end{figure}

We compare output histograms to noiseless theoretical histograms, all corresponding to 300 shots. As an illustrative example, we select an experiment employing GHZ state on 4 qubits, which is often considered as the most sensitive state to noise. In Fig.~\ref{noise}, parts (a) and (b) correspond to two different experiments, where in (a)  $\alpha=0.02\pi$ and $\gamma=\pi$, and in (b) $\alpha=0.165\pi$ and $\gamma=2\pi/5$. To simplify the presentation, we consider probabilities only for states  $|0000\rangle$ and $|0001\rangle$, which are the only states with non-zero probability in the absence of noise (In  Fig.~\ref{noise} this is illustrated as `Theory, noiseless').


We introduce independent single-qubit Pauli errors, according to the above noise model. The results are depicted in Fig.~\ref{noise} as `Theory, noisy'. We apply the same error model across all qubits, assuming errors within experimentally measured ranges. With the considered noise model, one can see a good match between noisy theoretical prediction (Fig.~\ref{noise}, `Theory, noisy') and experimental results (Fig.~\ref{noise}, `Experiment'). We compute the 1-norm of the probability vector space difference between the experiment and noisy simulation to demonstrate the agreement for the rest of the binary state subspace. The measured values do not exceed 0.05.

\begin{figure}[h]
\centering
\includegraphics[width=0.4\textwidth]{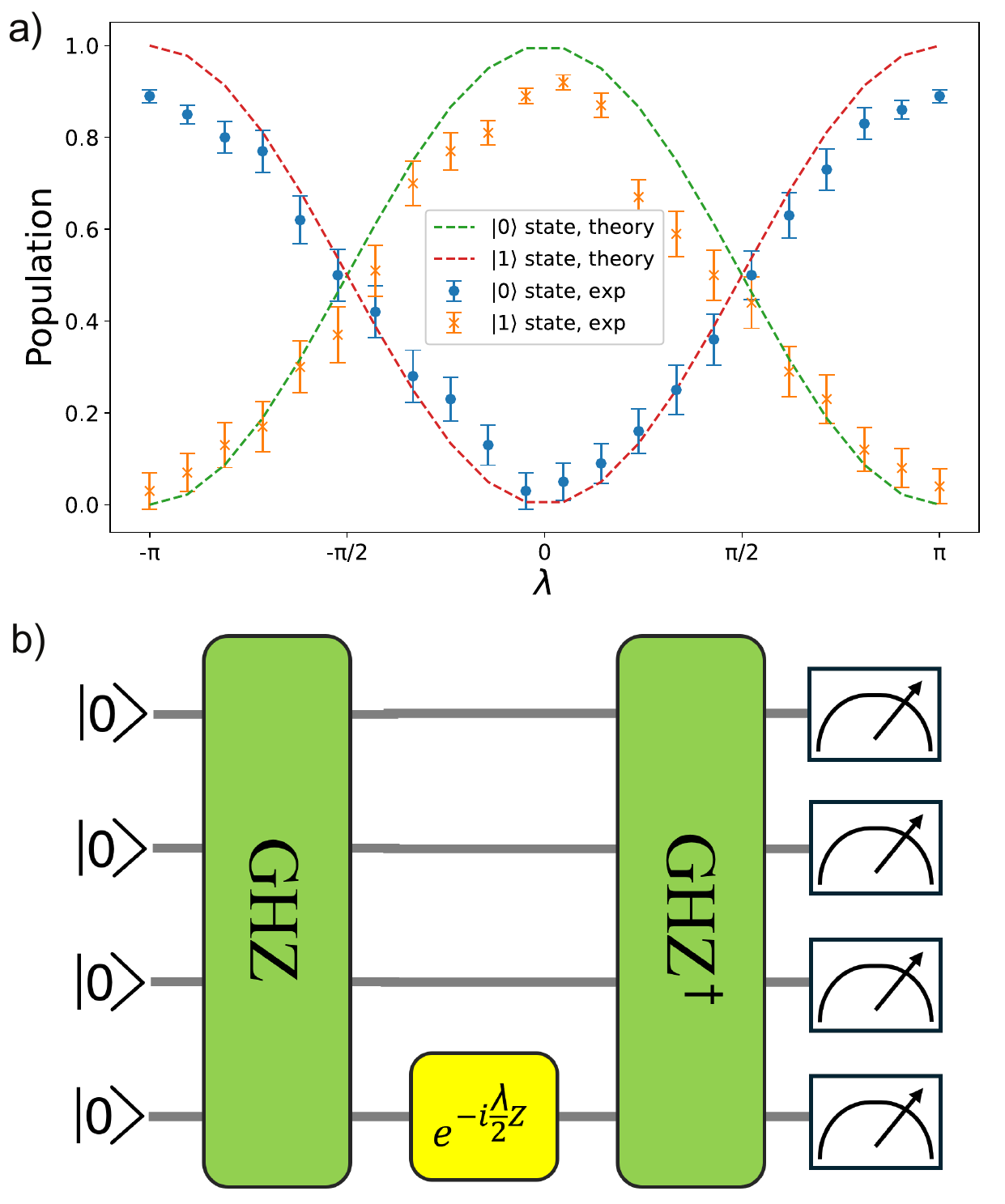}
\caption{\textbf{GHZ parity fringe benchmarking on four qubits.} a) GHZ parity fringe for the subset of four qubits in a seven-ion linear chain. 
We use the notation $|\textbf{0}\rangle=|0\rangle^{\otimes 4}$ and $|\textbf{1}\rangle=|1\rangle^{\otimes 4}$.  `Theory' curves represent parity fringe without noise or statistical error. The error bars on the figure represent two standard deviations ($2\sigma$) determined by the bootstrapping method discussed below. b) Schematic circuit for GHZ parity measurement. We vary the phase $\lambda$ to observe a parity fringe.}\label{GHZ_parity}
\end{figure}

\begin{figure}{\includegraphics[scale=.4] {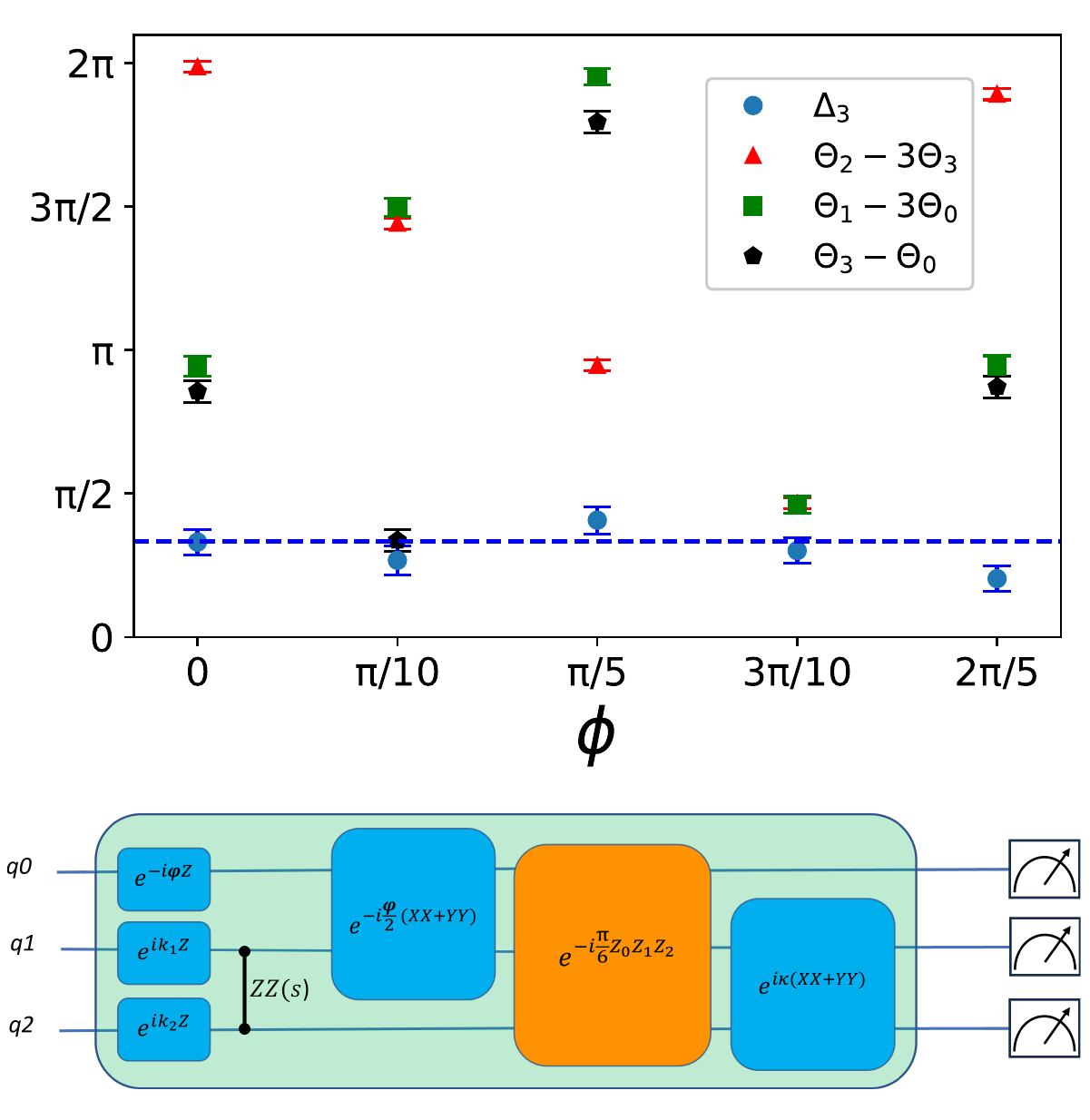}}\centering\caption{\textbf{Independence of $\Delta_{3}$ from one- and two-body interactions.} 
For the unitary realized in the bottom circuit, we measure the phase $\Delta_3$. The horizontal axis is $\phi$ that determines the phase of a one- and two-qubit gates in the circuit.  While all three phases 
$\theta_{2}-3 \theta_3$ (red), $\theta_1-3\theta_{0}$ (green), $\theta_3-\theta_0$ vary non-trivially with $\phi$, as predicted by our theory, the resulting total phase $\Delta_{3}$ (blue) remains approximately unchanged and consistent with the theoretical prediction of Eq.(\ref{Eq:Thm}), namely  $\Delta_3=-8\alpha_3= {\pi/3} \pmod {2\pi}$, within the error margins. The gate $ZZ(s)$ denotes $\exp(-\i \frac{s}{2} Z\otimes Z)$. In this experiment, $k_{1}=\pi/16$, $k_{2}= 3\pi/16$, $s=4\pi/5$, $\kappa=\pi/6$. } 
 \label{Fig:1_and_2_body}
\end{figure}

\noindent\textbf{GHZ parity benchmark:} We conduct a GHZ interference experiment to evaluate the performance of our system. When utilizing GHZ states, errors in individual qubits have a cascading effect,  impacting the overall performance. Hence, such experiments provide a powerful method for benchmarking the system.  Fig.~\ref{GHZ_parity}~(b) presents a schematic of the circuit employed for these measurements, and Fig.~\ref{GHZ_parity}~(a) displays the corresponding GHZ parity fringe observed across four qubits. We perform regular system calibrations to achieve this contrast level throughout the experiment.

\noindent\textbf{Insensitivity of  $\Delta_3$ with respect to 1- and 2- body Interactions:} In Fig.~\ref{Fig:1_and_2_body} we present the outcome of a 3-qubit experiment to verify that  $\Delta_3$ is indeed independent of one- and two-body interactions. The horizontal axis $\phi$ determines the phase of a single-qubit gate $\exp(-\i\phi Z_0)$ and two-qubit  $\exp(-\i\frac{\phi}{2} (X_0X_1+Y_0Y_1))$. We check that while the three phases $\theta_3-\theta_1$, $\theta_1-3\theta_0$ and $\theta_2-3\theta_3$ all vary non-trivially with  $\phi$, as expected from our theoretical results, the phase $\Delta_3= \theta_{n-1}-\theta_{1}-(n-2)\times (\theta_n-\theta_0)$ does not depend on $\phi$.

\begin{figure}[h]
\centering
\includegraphics[width=0.45\textwidth]{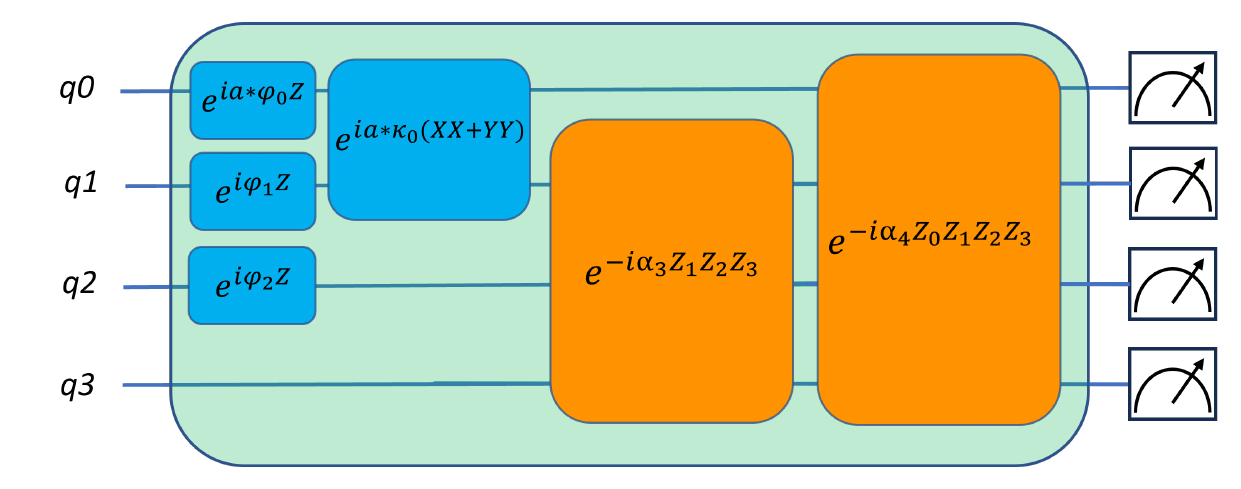}
\caption{\textbf{Circuits and corresponding parameters for $\Phi_{3}$ and $\Phi_{4}$ measurement.} In this circuit, the single-body $z$-rotation angles $\varphi_{0}$, $\varphi_{1}$, and $\varphi_{2}$ are respectively $-\pi/10$, $\pi/16$, $3\pi/16$, and $a$ is in the range from 0 to 10. The result of the measurement of $\Phi_3$ and $\Phi_4$ is presented in Fig.~\ref{fig_phi}.} \label{fig_phi_circuits}
\end{figure}

\noindent\textbf{Circuits for $\Phi_{3}$ and $\Phi_{4}$ measurement:}In Fig.~\ref{fig_phi} we presented the results of measuring  the phases $\Phi_{3}$ and $\Phi_{4}$ for the 4-qubit unitary realized by the circuit in Fig.~\ref{fig_phi_circuits}.   
 As expected from Eq.(\ref{def}), we observe that varying single- and two-qubit gates do not affect the measured $\Phi_{3}$ or $\Phi_{4}$ phases.

\section*{Acknowledgements}
This work is supported by the DOE Quantum Systems Accelerator (DE-FOA-0002253), the NSF STAQ Program (PHY-1818914), the NSF QLCI grant OMA-2120757, the Army Research Office
(W911NF-21-1-0005), as well as NSF grants Phy-2046195 and FET-2106448.




\bibliography{Ref_zhukas,Ref_2020,main2024,main}

\onecolumngrid
\newpage
\appendix

\newpage

\onecolumngrid

\onecolumngrid

\newpage

\maketitle
\vspace{-5in}
\begin{center}

\Large{Supplementary Material:\\ $ $ \\  Observation of the symmetry-protected signature of 3-body interactions }
\end{center}
\appendix

\section{Properties of phase $\Delta_3$ (Proof of Eq.(\ref{gen}))}\label{App:A}

 In this Appendix, we prove Eq.(\ref{gen}). To prove this equation  it is useful to remember some properties of operators 
\be\nonumber
C_l=\sum_{i_1<i_2<\cdots< i_l} Z_{i_1}\cdots   Z_{i_l}=\sum_{m=0}^n c_l(m)\  \Pi_m\ .
\ee
where the eigenvalues
\be\label{clcl}
 c_l(m)=\sum_{s=0}^l (-1)^s  {{m}\choose{s}} {{n-m}\choose{l-s}}\ \ \ \ \ \ \ :\ m=0,\cdots, n\ ,
\ee
are all integers. It can be easily seen that  $X^{\otimes n} \Pi_m X^{\otimes n}  =\Pi_{n-m}$, 
i.e., under applying Pauli $X$ on all qubits the $m$-excitation sector is transformed to  $m$-hole sector. Furthermore, since $C_l$ is decomposed as a sum of tensor products of $l$ Pauli $Z$ operators, we find $
X^{\otimes n}C_l X^{\otimes n}  =(-1)^l C_l$.   
This, in particular, implies 
\be\label{jkd}
c_l(m)=\frac{\Tr(C_l \Pi_m)}{\Tr(\Pi_m)}=\frac{\Tr(X^{\otimes n}C_l \Pi_m X^{\otimes n})}{\Tr(X^{\otimes n}\Pi_mX^{\otimes n})}= (-1)^l\frac{\Tr(C_l \Pi_{n-m})}{\Tr(\Pi_{n-m})}=(-1)^l\times  c_l(n-m)\ .
\ee
Operators $C_l: l=0,\cdots, n$ are linearly independent and satisfy the orthogonality relation
\be
\Tr(C_l C_{l'})=\delta_{l,l'}\times  \Tr(C^2_l)=2^n \times {{n}\choose{l}}\ .
\ee
It follows that the subspace spanned by $\{\Pi_m: m=0,\cdots, n\}$ is equal to the subspace spanned by operators $\{C_l: l=0,\cdots,n \}$. This subspace has dimension $n+1$. 
This is indeed the subspace of operators that commute with U(1)-invariant Hamiltonians.

Any operator  $H$ has a decomposition as
 \be\label{art}
H=H_\perp+\sum^{n}_{l=0} \frac{\Tr(HC_l)}{\Tr(C_l^2)} C_l\ , 
\ee
with the property that
\be\label{gh4}
\Tr(H_\perp\Pi_m)=\Tr(H_\perp C_l)=0\ \ \ \ \  : m,l=0,\cdots, n\ .
\ee

Applying this decomposition to operator $F_3=(n-2)(\Pi_0-\Pi_n)- (\Pi_1-\Pi_{n-1})$, we find that
 \be\label{art4}
F_3=\sum^{n}_{l=0} \frac{\Tr(F_3 C_l)}{\Tr(C_l^2)} C_l\ . 
\ee
Using the facts that $X^{\otimes n} F_3 X^{\otimes n}=-F_3$ and $X^{\otimes n} C_l X^{\otimes n}=(-1)^l C_l$, we find that 
\be
\Tr(F_3 C_l)=\Tr([X^{\otimes n}F_3 X^{\otimes n}] [X^{\otimes n}C_l X^{\otimes n}])=-(-1)^l \Tr(F_3 C_l)\ ,   
\ee
which implies $\Tr(F_3 C_l)=0$
for even $l$.

On the other hand, for odd $l$, Eq.(\ref{jkd}) we find
\be
\text{Odd } l:\ \ \Tr(F_3 C_l)=2(n-2)\times c_l(0)-2n\times c_l(1)\ .
\ee
Next, applying the formula for $c_l(m)$ in Eq.(\ref{clcl}) we find
\begin{align}
\Tr(F_3 C_l)&=2(n-2){{n}\choose{l}}- 2n {{n-1}\choose{l}}+ 2 n {{n-1}\choose{l-1}}\\
&=2\Big[(n-2)\frac{n!}{(n-l)! \times l!}- n \frac{(n-1)!}{(n-l-1)! \times l!}+n \frac{(n-1)!}{(n-l)! \times (l-1)!} \Big]\\
&=\frac{2\times(n-1)!}{(n-l-1)! \times l!} \times (\frac{n(n-2)+n l}{n-l}-n)\\
&=\frac{2\times(n-1)!}{(n-l-1)! \times l!} \times (\frac{2 n\times (l-1)}{n-l})\\ &=4(l-1)\times {{n}\choose{l}}\ . 
\end{align}
Putting this in Eq.(\ref{art4}) we find
\be
F_3=\sum^{n}_{l=0} \frac{\Tr(F_3 C_l)}{\Tr(C_l^2)} C_l= \sum_{l:\text{odd}}  4(l-1)\times {{n}\choose{l}} \frac{1}{2^n \times {{n}\choose{l}}} C_l=\frac{4}{2^n}\sum_{l:\text{odd}}  (l-1)  C_l\ .
\ee
This implies that
\begin{align}
\Delta_3&= -\int_0^T\hspace{-2mm} dt\  \  \Tr[H(t) F_3]\\ 
&=(n-2)\times (\theta_0-\theta_n)- (\theta_1-\theta_{n-1})\\ &=-\frac{4}{2^n}\sum_{l:\text{odd}}  (l-1)  \int_0^T\hspace{-2mm} dt\  \  \Tr[H(t) C_l]\pmod {2\pi},\nonumber
\end{align}
and  completes the proof of Eq.(\ref{gen}).

\color{black}

\section{An alternative approach for detecting 3-body interactions (Properties of phase $\beta_k$ and operator $B_k$)}\label{App:B}

We prove that for all $l,r\le n$, it holds that
\be\label{formula}
\sum_{l=0}^n  \Tr(B_r C_l)\times  (i\tan\theta)^l \ ={{n}\choose{r}} (2i)^r \frac{e^{i\theta (n-r)}}{ (\cos\theta)^{n-r} }\times (\tan\theta)^r\  \ \  \ \ \ \ \  \ \ \  :  -\pi< \theta\le \pi\ .
\ee
The proof is presented at the end of this section.  For $\theta=0$, the left-hand side of this equation becomes 
\be
\Tr(B_r)={{n}\choose{r}} (2i)^r  \delta_{r, 0}=\delta_{r, 0}\ ,
\ee
which means except $B_0=|0\rangle\langle 0|^{\otimes n}$ the rest of $\{B_r\}$ are traceless.

For $|\theta|\ll 1$, the right-hand side of Eq.(\ref{formula}) is equal to
\be
{{n}\choose{r}}  (2i)^{r}\times \theta^r+\mathcal{O}(\theta^{r+1})\ .
\ee
The left-hand side, on the other hand, becomes
\be
 \Tr(B_r C_{l_{\text{min}}})\times  (i\theta)^{l_{\text{min}}}+\mathcal{O}(\theta^{l_{\text{min}}+1})\ ,
\ee
where $l_{\text{min}}$ is the smallest value of $l$ for which $\Tr(C_l B_r)$ is non-zero.
Comparing the two sides, we can  immediately see that 
\bes\label{fff}
\begin{align}
\Tr(B_r C_l)&=0\ \ \ \ \ \  : l<r\\
\Tr(B_r C_r)&=2^{r}  {{n}\choose{r}} \ . 
\end{align}
\ees
Next, using this identity we prove lemma \ref{lem2}.

It is also worth noting that by taking derivatives with respect to $\theta$ of both sides of Eq.(\ref{formula}), at $\theta=0$, this equation implies
\be
\Tr(B_r C_l)=\frac{(-\i)^l (2\i)^r}{l!} {{n}\choose{r}}\times 
\frac{d^l}{d\theta^{l}}\Big[e^{\i\theta (n-r)}  \frac{(\sin\theta)^r}{ (\cos\theta)^{n}}\Big]_{\theta=0}\ .
\ee

\subsection{Proof of lemma \ref{lem2}}

First, using Eq.(\ref{art}) we find
\be
\Tr(B_r H )=\Tr(B_r H_\perp)+\sum_{l=0}^n \Tr(C_l B_r) \frac{\Tr(C_lH)}{\Tr(C_l^2)}\ .
\ee
Since $B_r$ is a linear combination of $\{\Pi_m\}$ operators, Eq.(\ref{gh4}) implies $\Tr(H_\perp B_r)=0$. Therefore, we find 
\begin{align}
\Tr(H B_r)&= \sum_{l=0}^n \Tr(C_l B_r) \frac{\Tr(HC_l)}{\Tr(C_l^2)}\\ &=\sum_{l=0}^k \Tr(C_l B_r) \frac{\Tr(HC_l)}{\Tr(C_l^2)}\ ,
\end{align}
where to get the second line we have used the assumption that $H$ can be written as a sum of $k$-local terms, and therefore it is orthogonal to operator $C_l$ for $l>k$. Applying Eq.(\ref{fff}), we find that for $r>k$, $\Tr(HB_r)=0$, and for $r=k$
\begin{align}
\Tr(H B_k) &=\sum_{l=0}^k \Tr(C_l B_k) \frac{\Tr(HC_l)}{\Tr(C_l^2)}\\ 
&= \Tr(C_k B_k) \frac{\Tr(HC_k)}{\Tr(C_k^2)}\\ &= \frac{2^{k}  {{n}\choose{k}}}{2^{n}  {{n}\choose{k}}} \Tr(HC_k)\\ &= 2^{k-n}\times  \Tr(HC_k)\ .
\end{align}
This proves lemma \ref{lem2}.

\subsection{Proof of Eq.(\ref{formula})}
First, using the Euler formula
$e^{i\theta Z}=\cos\theta I+i\sin\theta\ Z$, we find \begin{align}
 (e^{i\theta Z})^{\otimes n}&=(\cos\theta I+i\sin\theta Z)^{\otimes n}=\sum_{l=0}^n   (\cos\theta)^{n-l} (i\sin\theta)^l    \sum_{j_1< \cdots < j_l}Z_{j_1}\cdots Z_{j_l}  \\ &=\sum_{l=0}^n   (\cos\theta)^{n-l} (i\sin\theta)^l \ C_l\ \\ &=(\cos\theta)^{n} \sum_{l=0}^n  (i\tan\theta)^l \ C_l\ .
\end{align}
This implies that
\be\label{mnm}
\Tr((e^{i\theta Z})^{\otimes n} B_k)=\sum_{l=0}^n   (\cos\theta)^{n-l} (i\sin\theta)^l \ \Tr(B_k C_l)\ .
\ee
Next, we use the fact that
\be
 (e^{i\theta Z})^{\otimes n}=\sum_{m=0}^n e^{i\theta (n-2m)} \Pi_m\ ,
\ee
and 
\be
\Tr(\Pi_m)={{n}\choose{m}}\ .
\ee
This implies
\begin{align}
\Tr(B_k (e^{i\theta Z})^{\otimes n})&=\sum_{m=0}^k (-1)^m  \frac{(n-m)!}{(n-k)! (k-m)!}\times \frac{n!}{(n-m)! m!}  e^{i\theta (n-2m)}\\&=\frac{n!}{(n-k)!}\sum_{m=0}^k (-1)^m  \frac{1}{m! (k-m)!} e^{i\theta (n-2m)}\\&=\frac{n! e^{i\theta n}}{(n-k)! k!}\sum_{m=0}^k (-1)^m  \frac{k!}{m! (k-m)!} e^{-i\theta 2 m}\\&= e^{i\theta n} {{n}\choose{k}}\sum_{m=0}^k (-1)^m   {{k}\choose{m}} e^{-i\theta 2m}
\\&=  {{n}\choose{k}} e^{i\theta n}\times (1-e^{-i2\theta})^k\ .
\end{align}
In summary, we found
\begin{align}
\Tr(B_k (e^{i\theta Z})^{\otimes n})&=  {{n}\choose{k}} e^{i\theta n}\times (1-e^{-i2\theta})^k\\ &=  {{n}\choose{k}} e^{i\theta (n-k)}\times (e^{i\theta}-e^{-i\theta})^k\\ &= {{n}\choose{k}} (2i)^k e^{i\theta (n-k)}\times (\sin\theta)^k\ .
\end{align}
Comparing this with Eq.(\ref{mnm}) ,we find
\be
(\cos\theta)^{n} \sum_{l=0}^n  (i\tan\theta)^l \ \Tr(B_kC_l)={{n}\choose{k}} (2i)^k e^{i\theta (n-k)}\times (\sin\theta)^k\ ,
\ee
or, equivalently, 
\be
\sum_{l=0}^n  (i\tan\theta)^l \ \Tr(B_kC_l)={{n}\choose{k}} (2i)^k \frac{e^{i\theta (n-k)}}{ (\cos\theta)^{n-k} }\times (\tan\theta)^k\ .
\ee

\end{document}